# Label-Free Microrefractometry of Interfacial Processes Using Fluorescent Smart Coverslips

Hodaya Klimovsky,[1,2,3] Amitay Ginsberg,[1,2] Dmytro Ohorodniichuk[1,2], Maria Shehadeh,[1,2] Ilya Olevsko[1,2,3] Gerardo Byk,[2, ID] Martin Oheim,[3,4,ID,*] and Adi Salomon[1,2,3,ID,*,✉]

[1] Institute of Nanotechnology and Advanced Materials (BINA), Ramat-Gan, Israel;

[2] Department of Chemistry, Bar-Ilan University, Ramat-Gan, Israel;

[3] The NANOSCALE consortium, https://sppin.nanoscale.fr;

[4] Université Paris Cité, CNRS, Saint-Pères Paris Institute for the Neurosciences, Paris F-75006, France.

[ID]ORCID IDs: orcid.org/0000-0002-7159-9837 (GB), orcid.org/0000-0001-8139-167X (MO), orcid.org/0000-0002-5643-0478 (AS)

[*] Shared senior authorship

[✉] Corresponding author at adi.salomon@biu.ac.il

CONFLICT OF INTEREST STATEMENT

The authors declare no conflict of interest.

## AUTHOR CONTRIBUTIONS

A.S and M.O. supervised, conceived, and designed this research. G.B. supervised NBE synthesis. M.S. and D.O. synthesized NBE. H.K., M.S. and I.O. fabricated samples. H.K. and A.G. performed measurements and analysis. M.O. conceived and designed the microscope, A.S. and M.O. contributed to measurements and analysis. H.K. and A.G. programmed the image-analysis tools. All authors contributed to the interpretation of data. A.S and MO wrote the manuscript with contributions from all authors.

## DATA AVAILABILITY STATEMENT

The data that support the findings of this study and materials are available from the corresponding author upon reasonable request.

## SIGNIFICANCE

The refractive index (RI) quantifies chemical and physical changes in a medium, but fast refractometry at interfaces or in tiny volumes remains challenging. We introduce "smart" coverslips that convert subtle RI changes or nanoscale variations in transparent film thickness into measurable modifications of the radiation pattern emitted by a thin fluorescent layer deposited on a standard coverslip. By transforming a passive substrate into an active optical element, our method enables precise, label-free supercritical-angle fluorescence refractometry on a conventional fluorescence microscope, detecting $10^{-3}$ RI changes at 50 Hz, and opening new opportunities for studying surface reactions, biofilm growth, polymerization and membrane or organelle dynamics in living cells.

(107 words)

## ABSTRACT

Molecular dipoles near interfaces emit highly directional radiation due to near-field interactions, making surface-bound fluorophores sensitive probes of local physicochemical changes. We introduce "smart" coverslips, stably coated with uniform, brightly fluorescent nanobead films, that exploit refractive-index-dependent emission shifts for sensitive micro-refractometry in small volumes. Supercritical-angle fluorescence refractometry uses single back-focal-plane images to allow us real-time RI sensing and nanometric thin-film height measurements without the need for multi-angle or multi-wavelength acquisition. Our fast, label-free, and non-invasive approach allows measurements of thin-film properties and monitoring of interfacial dynamics on a standard inverted microscope and is broadly applicable to nano-biophotonics, chemical sensing, and *in-situ* materials analysis.

(98 words)

## INTRODUCTION

Fluorescence-based imaging and sensing are indispensable for studies in aqueous environments[1–5], but fluorescence intensity remains difficult to relate to quantitative physical variables, particularly near interfaces[6–8]. By contrast, the refractive index (RI) is a fundamental material parameter that directly reflects local polarizability and reports chemical and structural variations at interfaces[9–12]. While conventional refractometry accurately probes the RI of bulk media, resolving nanoscale RI gradients and interfacial order remains experimentally challenging and often model dependent. This is problematic, as the sub-wavelength regime (<100 nm), effective RI can deviate strongly from its bulk value, with spatial and temporal fluctuations encoding adsorption, surface reactions, and molecular assembly[9,13,14]. Existing approaches for measuring near-interface RI are slow, experimentally complex, or incompatible with standard optical microscopy in aqueous environments. Consequently, a fast and broadly accessible method for probing near-interface RI dynamics in real-time remains a major goal for interfacial science.

When fluorophores are located within ~100 nm of an interface, part of their otherwise undetected near-field emission couples to propagating modes, producing supercritical-angle fluorescence (SAF)[15–18]. The strength of this coupling depends on the fluorophore's axial position[19], orientation[6], and the local RI, making surface-bound emitters sensitive probes of interfacial RI variations[7,11,20] with a surface selectivity similar to that of total internal reflection fluorescence (TIRF) [1], but without the need for specialized excitation schemes[21]. Instead, SAF can be collected on standard inverted microscopes[22] using high-numerical aperture (NA) oil-immersion objectives and separated from undercritical-angle fluorescence in the back focal plane of the objective[16,23]. However, earlier reports of SAF-based refractometry[7] relied on fluorescent labeling of the sample or analyte, required long integration times, limiting its applicability for dynamic systems[7,11].

Here, we present a label-free approach for SAF-based refractometry at the micro- and nanoscale that makes use of *smart coverslips*[24]. Fluorescence is not generated in the sample but by a thin, transparent, and homogeneous nanobead-emitter (NBE) film stably deposited on a standard glass coverslip. This bright and uniform layer produces predictable and reproducible radiation patterns that are modulated by the local physico-chemical environment above the NBE layer, enabling RI detection of non-fluorescent samples. Importantly, our *smart coverslips* remain fully transparent and integrate seamlessly into standard fluorescence imaging and spectroscopy workflows.

Using calibrated sucrose solutions, we first demonstrate bulk RI measurements, with a sensitivity of ~$10^{-3}$ validated on an Abbe refractometer. We then apply our method to determine the *effective* RI of thin, transparent polymer films with thicknesses from a few nanometers (~λ/16) to several micrometers. We show that sensitivity is highest for film thicknesses from a few nm to ~λ/2 nm, consistent with the expected decay length of the dipole's near field. Of note, SAF refractometry on smart coverslips can distinguish thin films varying only by a few nm thickness. Finally, because a single BFP image suffices to recover RI or film thickness, our approach permits real-time tracking of rapid interfacial processes, like polymerization, surface reactions, biofilm growth, or liquid-gas phase transitions, the later of which we demonstrate here as a representative application.

(508 words)

## RESULTS

### Nanobead-emitter layers are an ideal, modular platform for 'smart coverslips'

We designed a flexible platform for near-interface label-free imaging and sensing, building on our recently developed nano-bead emitter (NBE) films[24]. NBEs are fluorophore-functionalized hydrogel nanoparticles that can be deposited as a monolayer on borosilicate (BK-7) coverslips using a modified layer-by-layer (LBL) technique. Their self-assembly leads to the formation of 8-nm thin, transparent, uniform, and exceptionally bright fluorescent films (**Fig. 1a**, *top*, and Fig. S1). When imaged with a high-numerical-aperture objective (measured $NA_{eff}$ = 1.465)[25] and Bertrand lens (to shift the image plane from the sample plane to the back focal plane, BFP), these films produced a highly, directional emission resulting from the conversion of the near-field emission of surface-proximal dipoles to supercritical angle fluorescence (SAF). On BFP images, SAF is detected as a bright ring surrounding the lower-intensity undercritical-angle fluorescence (UAF) (**Fig. 1a**, *middle*)[26,5,27,10]. As a result of the non-preferential dipole orientation, NBE films displayed a centrosymmetric emission pattern (Fig. S2), which permitted us azimuthal averaging of radial intensity profiles to yield virtually noise-free radiation patterns (see **Fig. 1a**, *bottom* for a polar-plot representation). The bright fluorescence of NBE layers was visible even to the naked eye, independent of the very fluorophore used and it did not saturate or bleach in the µW to low mW excitation regime (**Fig. 1b** and Fig. S3). NBE layers were uniform over large areas and displayed a sub-nm roughness ($R_q$ = 0.75 nm) (**Fig. 1c**)[24]. Collectively, these properties enabled wide-field imaging at high frame rates, with virtually indistinguishable radiation patterns down 1-ms exposure times (**Fig. 1d** and Fig. S4). The unique combination of controlled fluorophore height, isotropic orientation, high brightness, spatial uniformity, and molecular modularity makes our 'smart' coverslips a powerful toolbox for real-time interfacial sensing and imaging.

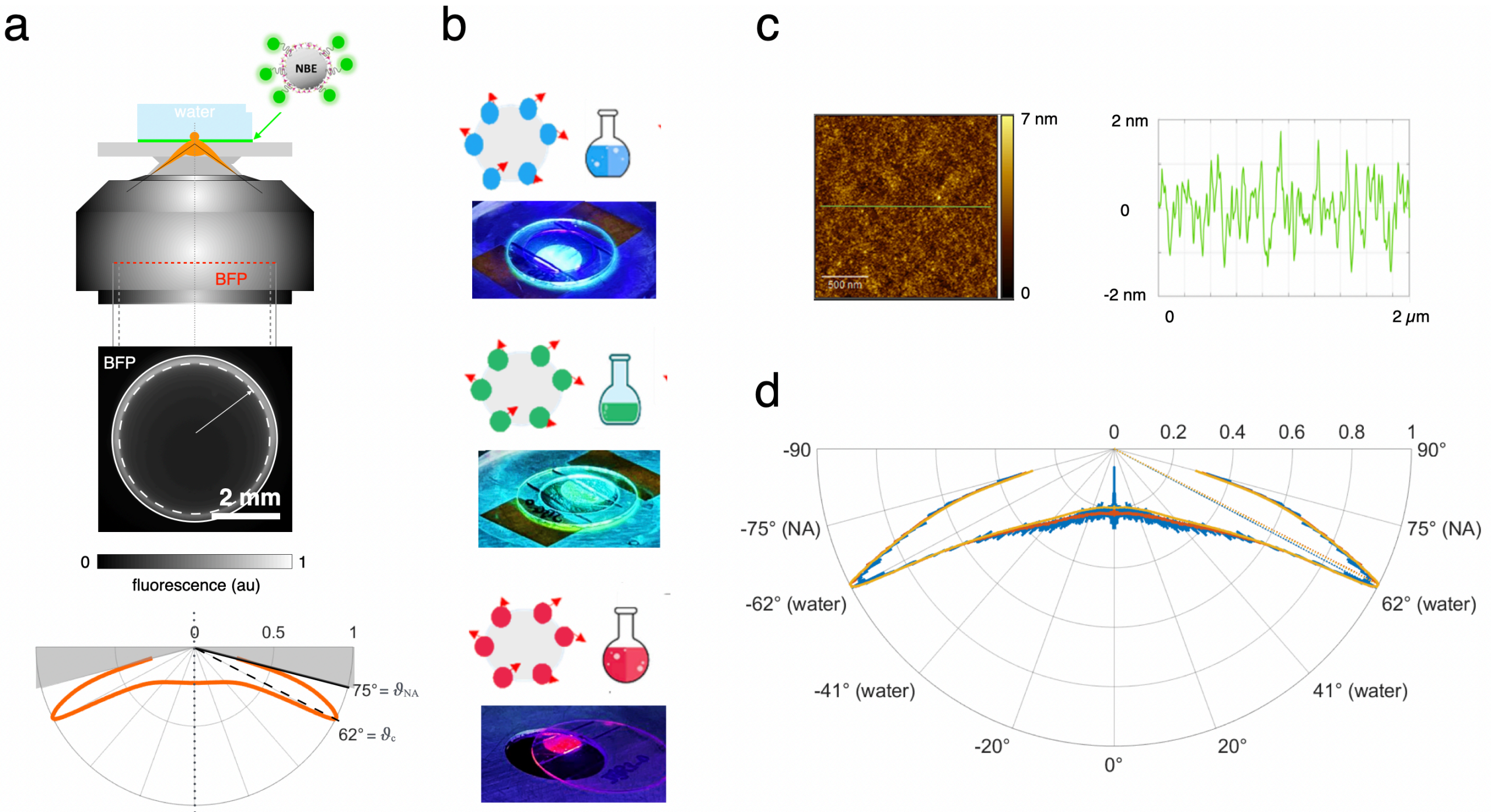

Fig.1 *Nanobead-emitters (NBEs) form bright, smooth and uniform fluorescent thin films on glass.* (a), experimental configuration: *top*, a high-NA objective captures an emission from a thin dye layer of NBEs (fluorophores encapsulated in a hydrogel nanoparticle, see *inset*). *Middle*, in the back-focal plane (BFP) of the objective light is sorted by the emission angle, not by the spatial position of the fluorophore. Near-surface fluorophores generate a strongly anisotropic radiation pattern with marked emission lobes at supercritical angles (SAF). *Bottom*, polar-plot representation of the azimuthally averaged BFP intensity profile. (b), examples of three different-color NBE layers (from *top* to *bottom*: ATTO425, ATTO488, ATTO590, respectively) illustrate the modularity of our NBE platform. Note the bright emission and transparency of the NBE layers. (c), *left*, atomic force microscopy (AFM) image of a NBE-ATTO488 film, and height profile (*right*) along the line indicated on the AFM image. Calculated RMS roughness was $R_q$ = 0.65 nm. (d), peak-intensity normalized and azimuthally averaged radiation patterns of a NBE-ATTO488 film topped with water for 1, 10 and 100-ms exposure times, in *blue*, *orange*, and *red*, respectively, reveal virtually identical radiation patterns.

**Smart coverslips allow precise and sensitive label-free refractometry**

The Brix scale (°Bx) expresses the mass percentage of sucrose in water, 1°Bx equal to 1 g of sucrose per 100 g of solution. The Brix scale is used as an index of total soluble solids in foods and beverages, and °Bx correlates with refractive index (RI). We prepared sucrose solutions with concentrations ranging from 0 to 40 °Bx, **Fig. 2a**, and measured their RI using an Abbe refractometer. Evanescent-wave excited BFP images of NBE-ATTO488 films topped with these solutions showed a gradual thinning of the SAF ring with increasing °Bx, **Fig. 2b**, as expected from the growing RI. The corresponding radiation patterns revealed a shift of the emission critical angle $\vartheta_c$ according to RI $\equiv$ NA = $n_2 \cdot \sin(\vartheta_c) = n_1$, **Fig. 2c** and **d**. Plotting the SAF- and Abbe-derived RI values revealed an excellent agreement, with unity slope and a minor offset of 0.013 RI units (Fig. S5), a notable improvement over what we observed earlier with native fluorophore layers using the same objective lens[28]. A corollary of this result is that the here used objective obeys Abbe-sine condition, i.e. that the peak position in the BFP can be easily related to the emission angles and hence the RI, even at the boundaries close to the NA.

Fig.2 *SAF-based microrefractometry.* (a), NBE-ATTO488 layers were overlayed with solutions of

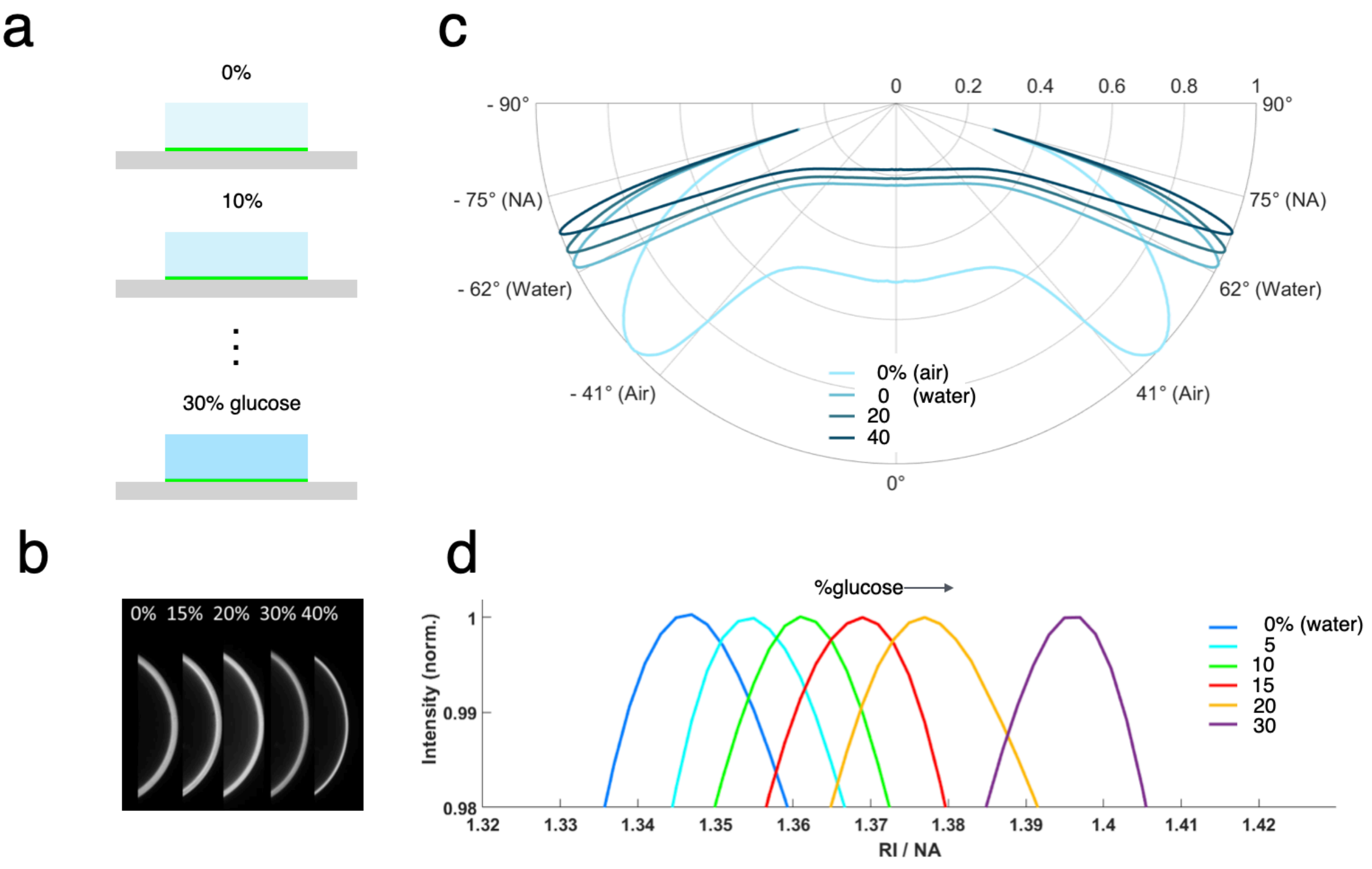

different sucrose content (in % of mass sucrose, °Bx). (b), examples of cropped BFP images at concentrations indicated showed a gradual thinning of the SAF ring with increasing °Bx. $t_{exp}$ = 1s. (c), azimuthally averaged radiation patterns confirm the systematic trend to higher-angle peak SAF emission with increasing RI. Radiation patterns of air (*light blue*) and water (*blue*) are shown for comparison. (d), crop on SAF peak region. Emission angles $\vartheta$ are expressed as NA values (or, equivalently, refractive index units, RI) according to NA = $n_2 \cdot \sin\vartheta_c \equiv n_1$.

Combining TIRF excitation with BFP imaging of NBE layers allowed us to directly measure the (bulk) RI of non-fluorescent, transparent solutions covering the NBE layer. SAF-based refractometry, combining BFP imaging, segmenting and measuring $\vartheta_c$, determined the local RI with ±0.002 resolution, limited by the BFP image pixel size (see discussion).

## Nanometric thickness measurements of transparent polymer thin films

The refractive index is fundamentally a bulk optical property, describing light propagation in macroscopically homogeneous media[9]. At micro- and nanoscales, however, this description breaks down as materials become optically heterogeneous. In nanostructured systems - such as photonic crystals, metamaterials, and nanoparticles - spatial inhomogeneities on the scale of, or below, the wavelength $\lambda$ lead to scattering, resonances, and field localization[13,14], giving rise to effective optical responses that can differ markedly from bulk behavior[13,10].
Similarly, the RI becomes ill-defined for thin dielectric coatings because the assumptions that make RI a meaningful material constant no longer hold at nm thicknesses. Thin coatings (typically <100 nm) are strongly influenced by the substrate and surrounding medium, so light no longer "sees" a uniform material.
To study how film thickness affects RI, we covered NBE-ATTO488 smart coverslips with MY-133-MC polymer. MY-133-MC is a moisture-cured, transparent and non-fluorescent (Fig. S6) material having a RI close to water (10, 25). As a control, we first drop-cast a ≈5 µm (~10·λ) polymer film and determined the precise thickness by profilometry (Fig. S7). The resulting NBE radiation pattern consistently exhibited the same SAF intensity peak and RI as bulk MY-133-MC (1.339 ± 0.002), close to that of water, 1.345 ± 0.002 (Fig. S8).

What happens for thinner films? We next prepared by spin coating MY-133-MC polymer films between roughly 30- to 150-nm thickness ($d \approx \lambda/16$ to $\approx \lambda/4$) on NBE-ATTO495 layers, **Fig. 3a**. Again, the precise thicknesses were obtained by stylus profilometry. In this regime of thin layers, the dipole's near-field evanescent emission should sample both the polymer film and the air above it, resulting in an effective refractive index ($RI_{eff}$) that scales with film thickness.

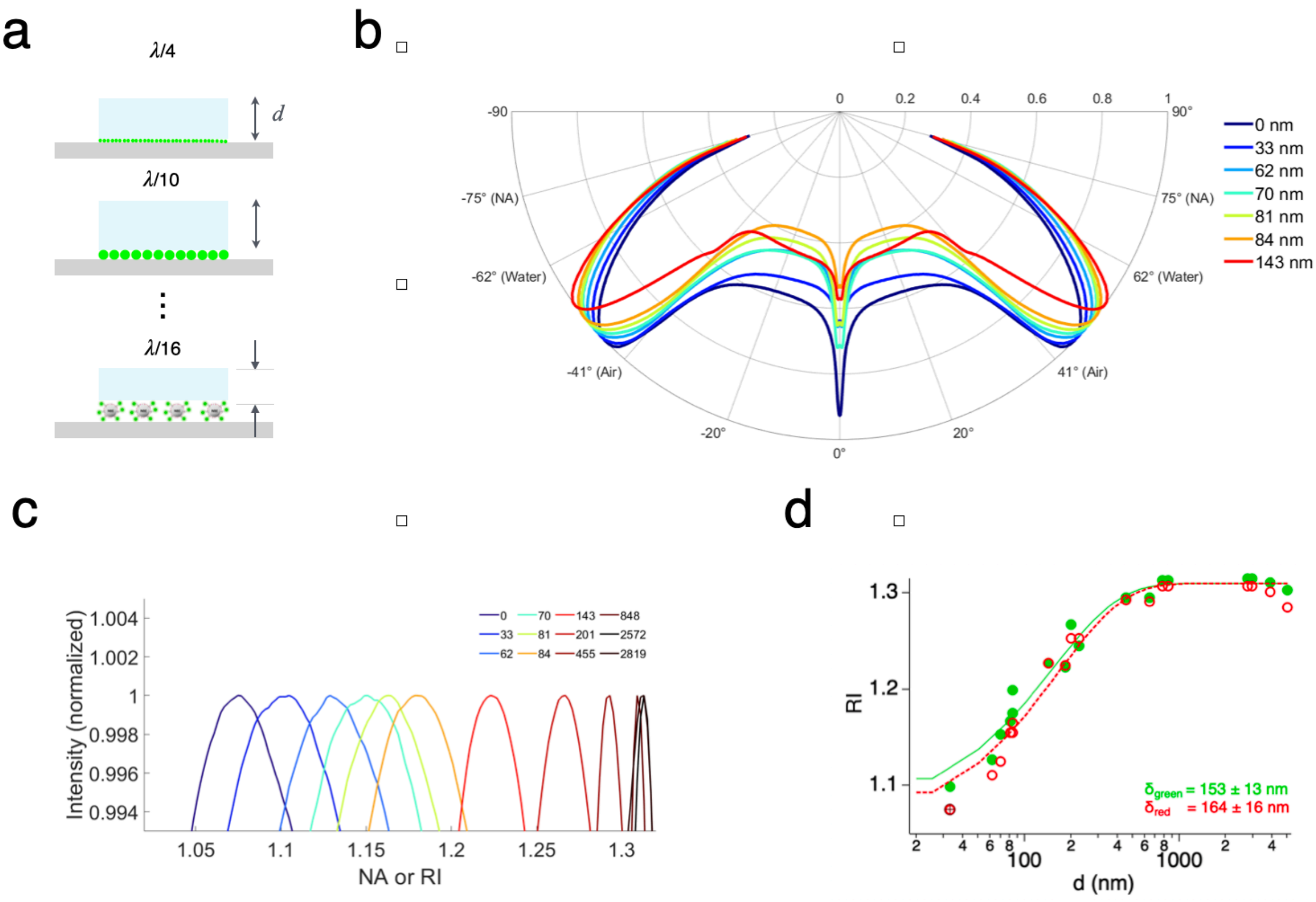

Fig. 3 *For sub-wavelength thin films of the same polymer material, the effective RI depends on the film thickness.* (a) illustration of the experiment: MY-133-MC polymer films of different nanometric thicknesses between 33 nm and 143 nm were prepared by spin coating on top of ATTO495 NBE layers. (b), measured radiation patterns, normalised to peak and color-coded according to film thickness $d$. $t_{exp}$ = 3-6 s (c), crop on the SAF peak region shows the monotonous displacement of the peak with film thickness and the resulting effective RI from close to air for very thin films to 1.31 for $d >> \lambda$. (d), apparent RI as a function of film thickness reveals a saturating function with the highest sensitivity for thicknesses below 200 nm. Graph shows combined independent measurements from different green- and red-emitting NBE-ATTO495 and NBE-ATTO490LS layers, as encoded by color. Through and dashed lines are fit of 1-exp(-d/$\delta$) with the data and revealed decay lengths of 153 nm and 164 nm for the green and red emitters, respectively.

Indeed, the measured emission patterns reported $RI_{eff}$ that is systematically shifted to higher values with increasing film thickness, consistent with a greater contribution of the polymer RI to $RI_{eff}$, **Fig.**

**3b**. We also systematically observed an emission maximum at 0 degree (along the optical axis), resulting from the illuminating laser spot in epi-fluorescence. For $\lambda/16$ thickness, the radiation pattern was barely distinguishable form that of air. Conversely, for thicker films the SAF peak maxima increasingly migrated to higher RI values, **Fig. 3c**. Also, those SAF peaks became progressively slimmer for higher RI values, as expected from Abbe's sine condition: at larger emission angles, more angular components are mapped into a given radial (NA or RI) interval in the BFP [29]. The sensitivity to film thickness is highest in the range of 100 to 400 nm, commensurable with the decay length of the NBE near field emission. On the plot of $RI_{eff}$ *vs.* film thickness, we see that $RI_{eff}$ converges to its bulk value for $d > 2\cdot\lambda$. We also compared the dependence of $RI_{eff}$ on film thickness $d$ for two different NBE layers having different emission spectra. When fitting a monoexponential $(1.318\cdot(1\text{-}\exp[-d/\delta]))$ with measured $RI_{eff}(d)$ data, the obtained decay lengths of 153±13 nm and 164±16 nm, scaled with the peak emissions of ATTO488 (530 nm) and ATTO490LS (620 nm), respectively, **Fig. 3d**.

Taken together, the data shown in figs. 2 and 3 demonstrate our ability to measure *effective* refractive indices of non-fluorescent solutions and thin films with high precision, by detecting minute surface-induced polarizability changes that modify the emission pattern of the fluorophores of our 'smart' coverslips. We can resolve film-thickness differences on the order of 5 nm, providing sufficient sensitivity to track crystallization, aggregate formation, or the onset of polymerization reactions. Our technique provides a robust measure of the *effective* RI with an accuracy of ±0.002 RI units and the highest sensitivity in the sub-$\lambda$ regime. Conversely, for materials of known RI, our measurements of an effective RI, can be directly translated to thickness measurements with nm precision. This is important for soft materials, in which thickness measurements are method dependent.

### SAF-microrefractometry permits real-time monitoring dynamic chemical processes

We next demonstrate that our approach can resolve interfacial transformations in real time. As a proof-of-concept, we monitored the evaporation of a hydrofluoroether (HFE), which exhibits a high vapor pressure at room temperature (22-23 °C) despite its elevated boiling point (≈130 °C)[30]. We deposited a 2-µl HFE droplet on a SiOx-protected NBE-ATTO488 smart surface and we recorded a BFP-image time series, (**Fig. 4a**, *top*). The radiation patterns extracted from these images evolved with time. At $t$ = 0, the NBE radiation pattern and the SAF peak matched the RI of liquid HFE (n≈1.294), **Fig. 4b**, *left*. By ~22 s, it had converged to the RI of air, marking complete solvent evaporation, **Fig. 4b**, *right*. Between these endpoints, the patterns progressed through two distinct regimes (see **Supporting Video S1**). Early in the process, the SAF peak exhibited no detectable shift while a second, minor hump emerged at lower RI, the intensity of which increased with time. Around 13 s, a second, clearly resolvable SAF band is seen with a peak at the RI of air, indicating the coexistence of liquid and vapor microdomains within the same field of view. An interesting feature is the dark band between the two emission peaks (**Fig. 4b**, *middle*), before the two-ring pattern collapsed and gave rise to the familiar, broad emission lobe of NBE-ATTO488 exposed to air (**Fig. 4c**).

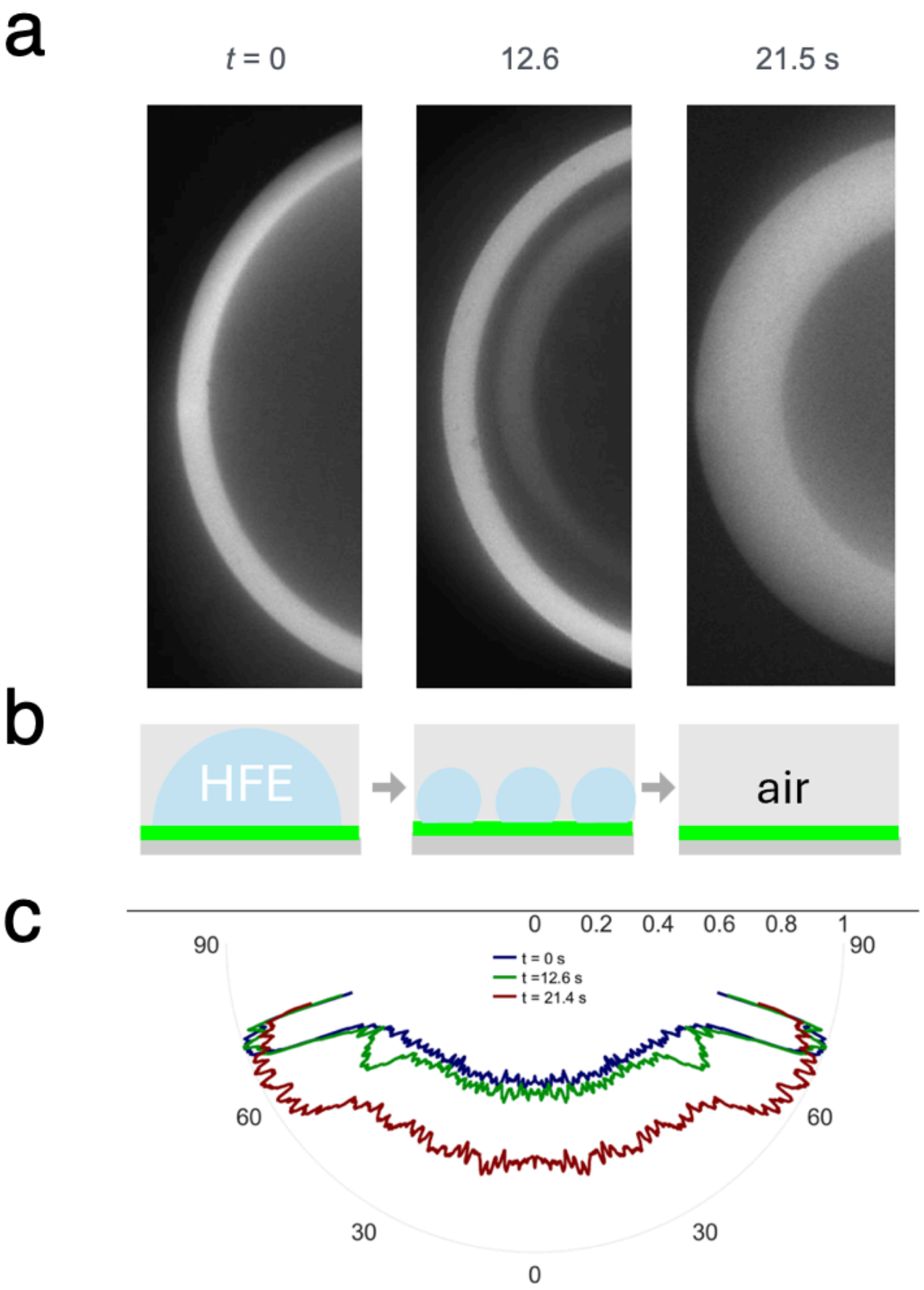

Fig. 4 *Monitoring solvent evaporation on smart coverslips.* (a), selected BFP images at times indicated of a 2-µl drop of Hydrofluoroether (HFE) on a NBE-ATTO488 layer ($t_{exp}$ = 200 ms, frame rate = 1.28 Hz). See Supporting Video S1 for the complete data set. (b), sketch of our interpretation of the observed radiation patterns. (c), radiation patterns extracted from the BFP images in (a) at the beginning (*blue*), during evaporation (*green*) and at the end of the experiment (*red*), respectively,

show an evolution from a pure HFE pattern through the coexistence of two states to a pure air pattern following the full evaporation. See main text and supporting information for details.

We can draw several interesting conclusions from these observations. First, the measured BFP radiations patterns allow us to clearly distinguish a continuous effective-RI regime (as shown in **Fig. 3**) from a genuine two-phase transition (**Fig. 4**). For multilayered environments, the NBE emitters 'sense' a stacked optical environment, and we observe a single SAF lobe that changes in width and shifts smoothly with film thickness $d$. In this case, the system can be described by $RI_{eff}$, and the SAF pattern evolves continuously, until the RI converges to the bulk value for film thicknesses around $2\lambda$ (as seen on **Fig. 3d**).

By contrast, during solvent evaporation **(Fig. 4)** we observe two separate rings on the BFP image (or equivalently two distinct SAF peaks): one associated with the liquid RI and one with the RI of the vapor/air interface. This indicates that the emitters within the field-of-view are simultaneously exposed to two distinct dielectric nano-environments - liquid and gas - rather than a single dipole sensing a mix of two RIs. This scenario is reminiscent of forbidden bands in metamaterials. Thus, the dark annulus between the two SAF rings could correspond to the range of forbidden emission angles (or transverse wavevectors $\boldsymbol{k_t}$ which are not allowed). These angles are unoccupied, because there is no stable layer having a mixed RI between that of the liquid and the vapor. That is, the angular gap in the BFP image reflecting a bimodal distribution of nano-environments, or in chemical terms, the co-existence of phases (Movie S1 and S2).

A second noteworthy feature is that the dual-ring pattern disappears abruptly once the liquid phase is gone, and the BFP collapses back to a single SAF ring, signaling a transition from a two-phase (liquid-vapor) state to a single dielectric phase. The evolution from liquid to gas thus cannot be represented as a linear combination of the two limiting emission patterns, confirming that evaporation proceeds via non-equilibrium pathways with transient phase coexistence, rather than through a simple, homogeneous thinning like that would be described by a single changing $RI_{eff}$ (**Fig. 3**).

We note that the kinetics of this process is further modulated by interfacial energetics: surface energy and contact angle determine the lateral distribution and curvature of the receding film, thereby governing the duration and extent of phase coexistence (data not shown)[31,32]. Overall, the observed two-phase behavior reveals a complex interplay between continuous de-wetting and lateral phase separation, reminiscent of "coffee-ring"–type evaporation[33,34]. The ability to resolve these mixed regimes with sub-second temporal resolution highlights the power of SAF micro-refractometry with smart coverslips for capturing dynamic interfacial transformations.

## DISCUSSION

We here introduced a SAF-based strategy for label-free micro- and nano-refractometry that can be implemented on any inverted microscope equipped with a sufficiently high-NA objective, requiring a Bertrand lens as the only add-on. Although our method uses far-field detection, it provides precise RI readout from a nanometric interfacial region by exploiting subtle modifications of the dipole radiation pattern, which originate from near-field coupling and are translated into predictable and measurable changes in the objective BFP. A key enabling element for this technique are our NBE-based *smart coverslips*. Unlike native fluorophores, NBE layers provide uniform, bright, stable fluorescence from isotropically oriented emitters yet they are axially confined to ~8 nm. Advantageously, the modular design of fluorophore-decorated hydrogel nanoparticles decouples fluorophore chemistry from film formation. A consequence is that our technique supports a wide range of dyes, and it yields spatially uniform signal levels over millimeter scales. No labeling of the sample or analyte is needed, and the signal is high enough to track dynamic nanoscale processes at the sub-diffraction scales described in[35].

In the current implementation the measured radiation patterns are averaged over the entire FOV, however, the high brightness of our NBE films allows for a scanning geometry with spot excitation, paving the way for spatially resolved RI measurements[36].

**Comparison with other techniques.** Several optical techniques can probe RI near interfaces, yet each faces its own practical limitations. Ellipsometry and reflectometry offer sub-nanometer sensitivity to thin-film thickness and optical constants, but they rely on model fitting and are largely restricted to homogeneous, static layers[37–39]. Near-field optical microscopies (e.g., SNOM) achieve nanometric spatial resolution, but they suffer from mechanical instability[40], small fields of view (FOV) and low throughput, as well as limited compatibility with liquid environments[41]. Interferometric scattering (iSCAT) and quantitative phase imaging (QPI) can retrieve RI variations with high temporal resolution, but they require complex reconstruction algorithms and careful in-situ calibration[42,43]. Surface plasmon resonance (SPR) provides exceptional RI sensitivity ($\Delta n \approx 10^{-6}$–$10^{-7}$) on a similar length-scale than our technique (~100–200 nm), but it relies on opaque metal-coated substrates, which hinder simultaneous optical imaging and spatial mapping [13]. Similar to our approach, variable-angle TIR measures RI values from the abrupt change in the reflected laser intensity near the (excitation) critical angle [44,27], but it requires the acquisition of a stack of images to sample the angles around $\vartheta_c$, slowing down acquisition and precluding monitoring of fast phenomena. In this context, NBE-enabled SAF-based micro- and nano-refractometry occupies a unique niche. Based on the analysis of a single BFP image, it simultaneously offers sensitivity, speed and large-field readout. With the use of *smart coverslips*, no sample labelling is required, but the technique remains fully compatible with fluorescence imaging and sensing in spectral channels not occupied by NBE emission.

**Near-field readout by far-field imaging.** We demonstrated that for transparent thin films, SAF refractometry reports an *effective* refractive index that scales with film thickness. Our method achieves sub-wavelength sensitivity by exploiting the RI-dependent coupling of evanescent dipole emission components to the dielectric interface, operating on a length-scale below approximately 400 nm. Measuring thin-film thickness becomes possible because the NBE emitters are stably confined to an ~8 nm layer and they are, on average, isotropically oriented, so that changes in the SAF pattern uniquely report on variations in the local dielectric environment and they are not confounded by fluorophore height or dipole orientation, as in sample-labeled approaches. The slight but measurable shift in the $RI_{eff}$ vs. thickness curve is a direct consequence of the dependence of decay the dipoles' near-field emission component on the emission wavelength, and - given the modular architecture of the NBE platform - can be used to fine-tune sensitivity and length-scale of the layer characterisation.

**Dynamic interfacial processes and evaporation-driven dewetting.** A final consequence of the brightness of our NBE layers is that *smart* coverslips enable time-resolved studies with high time-resolution. Our evaporation experiment illustrates their potential as powerful probe of non-equilibrium interfacial dynamics. By converting minute changes in local polarizability and film morphology into robust *k*-space signatures, our approach combines a nanophotonic readout with macroscopic interfacial processes. The fact that continuous thinning, liquid-vapor phase coexistence, and complete dewetting can all be distinguished optically on a standard microscope suggests that SAF microrefractometry can become a general tool for studying thin-film formation, evaporation-driven deposition, polymerization, and eventually more complex catalytic and biological interfacial transformations.

Evaporation-driven dewetting is more than a convenient test system; it is representative of a broad class of non-equilibrium processes at solid–liquid interfaces. Drying of droplets underpins coffee-ring formation, inkjet and spray printing, thin-film coating, and the preparation of biological and analytical samples. In all of these cases, the kinetics of film thinning, breakup, and contact-line recession govern the final distribution of solutes, nanoparticles, or polymers on the substrate. Our results show that SAF microrefractometry can resolve this transformation with sub-second temporal resolution and a refractive-index sensitivity of $\Delta n \approx 0.002$, while simultaneously distinguishing continuous thinning from lateral phase separation. This capability positions our smart coverslips as a general platform for studying deposition and pattern formation under real-world conditions and on relevant time-scales.

## EXPERIMENTAL PROCEDURES

*Smart coverslip fabrication.* #1 BK-7 glass coverslips (Menzel Gläser, Braunschweig, Germany) were cleaned in 1/100 Hellmanex/double-distilled-deionized water (DI water, 18.2 mΩ) and sonicated (20 min, 40 °C). They were then rinsed thoroughly with DI water, ethanol, and DI water again (30 s each). Substrates were dried with a $N_2$ flow and immediately used thereafter. NBE synthesis and deposition were done as is described in ref. 24. For HFE evaporation process, NBE layers were covered by $SiO_x$ thin film (~ 10 -15 nm) for protection and chemical stability. The resulted effective RI following the SiOx layer is shifted by 0.04. The smart coverslips were characterized prior to use to verify their homogeneity and smoothness, as described earlier [24].

*Polymer thin films.* We used MY-133-MC (MyPolymers, Ness Ziona, Israel), a moisture-cured (MC) transparent, non-fluorescent (Fig. S6) polymer, with a refractive index of 1.339 (at 488 nm). Due to its RI close to water, it is used in various bio-photonic applications, such as SPR bio-sensors or TIRF calibration samples[10,25,45];

*Polymer deposition.* MY-133-MC was directly applied onto our smart *coverslips*. Polymer solution was filtered, and thin films were spin-coated (WS-650, Laurell Tech Corp., North Wales, PA) and cured under ambient conditions (room temperature, 20–22 °C) during at least 6 h. Depending on the final desired thickness, My-133-MC polymer resin was applied either pure (undiluted) or diluted in HFE-7500[46].

*Layer characterisation.* Thin polymer layers were characterised as described earlier [10]. The layer thicknesses reported in the figure of this paper are based on measurements using stylus profilometry (DektakXT, Bruker) in at least 6 areas. As expected for soft matter, the measured thickness depended on the applied Stylus force, between 0.3 µN to 30 µN, with higher force applied resulting in thinner apparent thicknesses due to the indentation of the polymer layer. The plot of measured layer thickness $d$ against $log(F)$ revealed an almost mono exponential dependence of $d$ on $F$ (Supporting Fig. S7). We fitted, for each produced layer, a falling exponential with the measured $d(F)$ and chose the intersection of the curve fit with 1 µN force as its *effective* thickness. Positive and negative error bars were generated from the values at 0.5 and 1.5 µN, respectively.

*Atomic force microscopy*. AFM images were acquired on a Veeco Nanoscope. The AFM images are shown as pseudo colored height maps, from which we calculated the mean squared roughness, $R_q$, i.e., the quadratic mean of profile height deviations from the mean (Gwyddion).

*Epi-fluorescence measurements.* Emission spectra were acquired on an inverted microscope (IX83, Olympus) coupled to an IsoPlane SCT320 spectrophotometer (with a 600-nm blaze and 50 grooves per nm grating), and equipped with a PIXIS 1024 eXcelonTM charge-coupled device (CCD) camera (TeleDyn Princeton Instruments). Spectra were acquired upon 405/488 nm excitation with filters shown in Fig. S9. The spectrometer was calibrated with the Princeton Instruments IntelliCal calibration device. The excitation source was an Xe-arc lamp in conjunction with suitable excitation filters and dichroic mirrors. Exposure times ranged from 1 to 10 s. The presented emission spectra were average spectra derived from 1023-pixel rows, analyzed using MATLAB software.

*Combined TIR-SAF microscopy.* Fluorescent layers were imaged on a custom inverted microscope, assembled from optical bench components (see Fig. S10 for details). The optical path allowed for adjustment of the excitation beam angle as well as toggling between sample-plane (SP) and back-focal plane (BFP) imaging by toggling in and out a Bertrand lens. The beam of a 488-nm laser (Coherent Sapphire SF 488-50) was cleaned with a notch filter, attenuated with neutral density filters, spatially filtered, and expanded to ≈2″ diameter, resulting in illumination of the entire field of view. The beam was scanned by a pivoting mirror and focused on the BFP of a Plan-Apochromat ×100/NA1.46 Oil DIC M27objective (ZEISS). Images were captured using high-angle evanescent-wave (EW) or epifluorescence (EPI) excitation, with laser powers and exposure times as indicated. Fluorescence was collected through the same objective, filtered by long-pass filters (488LPXR,

RET493LP, ET520LP), and imaged on a back-illuminated sCMOS camera (PCO edge4.2bi). The laser, camera and all motorised setup components were controlled by a custom graphical user interface (GUI) described elsewhere [10]. The effective pixel size in the sample plane was 30 nm/px, allowing a two- to fourfold binning without loss of resolution. Swinging in the motorized Bertrand lens shifts the image plane to the objective's BFP, which on our microscope results in an effective BFP pixel size of 3.36 µm/px. For diffraction-limited BFP imaging, we can afford a BFP pixel size of 5 µm, meaning that either we oversample (for unbinned BFP images) or we slightly sub-sample (for 2×2 binning).

Laser, camera and filter spectra are compiled in the supporting material (Fig. S11).

*AI-based segmentation and azimuthal averaging*. SP and BFP fluorescence images were subtracted with their respective dark images. BFP images were segmented into SAF and UAF. Given the previously measured effective NA of the objective ($NA_{eff}$ = 1.461 ± 0.001, i.e., the outer boundary of the BFP image), its value serves as a reference for measuring the radius of the critical angle $\vartheta_c$ (inner ring, SAF/UAF transition) that is related to the sample RI, $n_1$, as $r = f \cdot n_2 \cdot \sin\vartheta_c = f \cdot n_1$.

As our NBE layers show no preferential dye orientation and generate radially symmetric BFP images (see Fig. S2) we azimuthally averaged radial intensity profiles for noise reduction. We first used the Segment Anything Model (SAM) implemented in MATLAB to find the BFP image center. Briefly, the user manually selects the BFP image region of interest (ROI), SAM then generates a binary image, which is fitted with a circle to determine center coordinates ($x_0$, $y_0$). Together with the profile length, ($x_0$, $y_0$) are used to trace a radial vector from the center to $r_{NA}$. After passing to radial coordinates, $(x, y) = [x_0 + r \cdot \cos(\vartheta), y_0 + r \cdot \sin(\vartheta)]$, the computed values are rounded, and the corresponding intensity for each $r$ is stored. The azimuthally averaged intensity profile is the mean intensity profile over all angles.

*Statistics.* All experiments were performed at least as triplicates of independent experiments. Figures show population mean ±SD. Error bars are shown if the error was bigger than the symbol size or line width. Non-normally distributed data are shown to appreciate the spread of points.

SUPPORTING INFORMATION

Supporting Information is available on the PNAS website.

ACKNOWLEDGEMENTS

The authors thank Mor Hale (BIU Engineering Department) and Dr Omer Shavit (SPPIN) for helpful discussions. This study was financed by a joint Franco-Israeli PHC Maimonide grant (CC-NanoCOQ, to M.O and A.S.), by the Centre National de la Recherche Scientifique (CNRS), University Paris Cité, and by the European Union (H2020 Eureka! Eurostars, "NANOSCALE" (E! 12848, https://nanoscale.sppin.fr, to M.O. and A.S.) The authors are grateful for mobility support from Franco-Israeli CNRS-IRN, "OMNI-TOOLS," and the France-Bio-Imaging large-scale national infra-structure initiative (FBI, ANR-10-INSB-04, Investments for the future).

SUPPORTING INFORMATION

# Label-Free Microrefractometry of Interfacial Processes Using Smart Coverslips

Hodaya Klimovsky, Amitay Ginsberg, Dmytro Ohorodniichuk, Maria Shehadeh, Ilya Olevsko Gerardo Byk, Martin Oheim*, and Adi Salomon*.

**Corresponding author:** adi.salomon@biu.ac.il

This PDF file includes:

- List of abbreviations
- Supporting text
- Supporting figures S1 to S11
- Supporting tables S1 to S2
- Legends for movies S1 to S2
- SI References

Other supporting materials for this manuscript include the following:

Movies S1 to S2

## LIST OF ABBREVIATIONS

| | | |
|---|---|---|
| AF | - | autofluorescence |
| AFM | - | atomic force microscopy |
| AI | - | artificial intelligence |
| BFP | - | back-focal plane |
| BK-7 | - | borosilicate (glass) |
| °Bx | - | degree of Brix (% mass/mass of sucrose in water) |
| CCD | - | charge-coupled device |
| CV | - | coefficient of variation |
| DI | - | deionized |
| EPI | - | epifluorescence |
| FA | - | full aperture |
| GUI | - | graphical user interface |
| HFE | - | hydrofluoroether |
| iSCAT | - | interferometric scattering |
| LBL | - | layer-by-layer |
| MC | - | moisture-cured |
| MIET | - | metal-induced energy transfer |
| MY-133-MC | - | MyPolymer, $n$ = 1.33, moisture-cured |
| NA | - | numerical aperture |
| NBE | - | nanobead emitter |
| PTFE | - | Polytetrafluoroethylene |
| QPI | - | quantitative phase imaging |
| RI | - | refractive index |
| RMS | - | root-mean square |
| SAF | - | supercritical angle fluorescence |
| SD | - | standard deviation |
| SNOM | - | scanning near-field optical microscopy |
| SPR | - | surface plasmon resonance |
| TIR(F) | - | total internal reflection (fluorescence) |
| UAF | - | undercritical angle fluorescence |
| VA-TIRF | - | variable-angle TIRF |

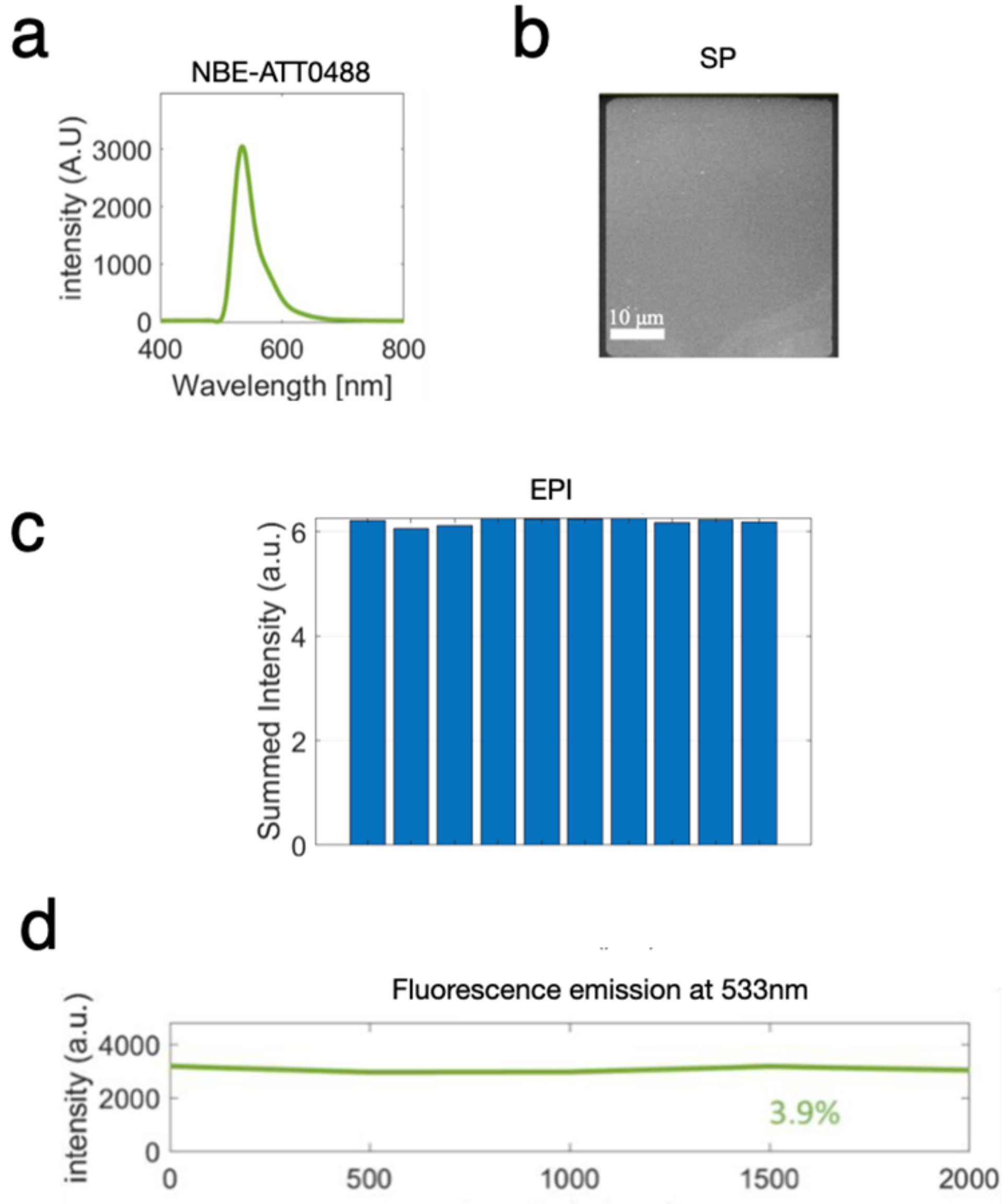

**Fig. S1.** *NBE layers are uniform over mm-distance.* (a) fluorescence spectrum of a NBE-ATTO488 layer of nanobead emitters (NBEs) upon 488-nm epifluorescence excitation (EPI). $\lambda_{em}^{(max)}$ = 533 nm, vs. 523 nm for the dye in aqueous solution (see Table S1). (b), EPI sample-plane (SP) image of the same sample. (c), comparison of integrated fluorescence measured upon EPI excitation in ten different regions of interest (ROIs) on the same NBE-ATTO488-coated coverslip show minimal intensity difference. (d), intensity profile along a 2-mm line from a lower-magnification image shows a coefficient of variation (CV = SD/mean) <4%, which - together with the AFM images in the main text - illustrates the exceptional flatness and uniformity at of our NBE layers various length-scales. See ref. [1] for a more complete characterization.

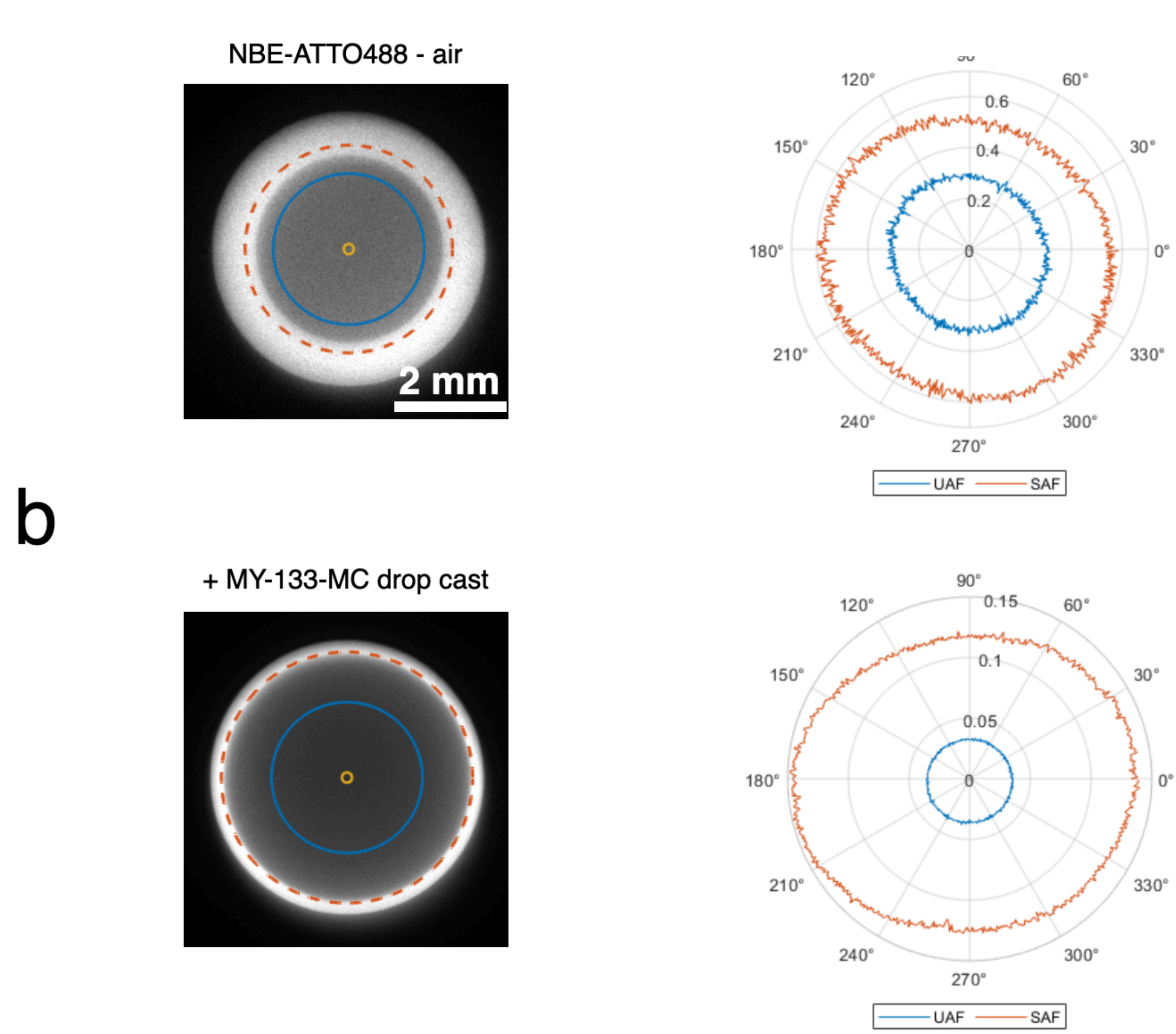

**Fig. S2.** *Emission patterns from NBE-films are largely centrosymmetric* NBE-ATTO488 was deposi-ted onto a plasma-cleaned and PDDA-coated BK-7 substrate and imaged upon evanescent-wave excitation (TIRF) at 488 nm. Images show, *left*, the BFP image and, *right*, the polar-plot representation of the azimuthal intensity profile shown in *blue* and *orange* for UAF and SAF zones, respectively. No protective silica layer was applied on top of the NBE layer in this experiment. $t_{exp}$ = 100 ms. (a), NBE-ATTO488/air interface. Overlayed on the BFP image are the center found with SAM, *yellow*, as well as circular regions used for plotting the intensity profiles at undercritical ($NA_{UAF}$ = 0.8), and supercritical angles ($NA_{SAF}$ = 1.1), respectively. (b), same, following the deposition of a µm-thick MY-133-MC polymer layer by drop cast. ($NA_{UAF}$ = 0.8 and $NA_{SAF}$ = 1.333, respectively). Note the near-perfect azimuthal isotropy (centrosymmetry) for UAF and a very high degree of symmetry for SAF emission in either case, justifying the azimuthal averaging of radiation patterns applied in this work.

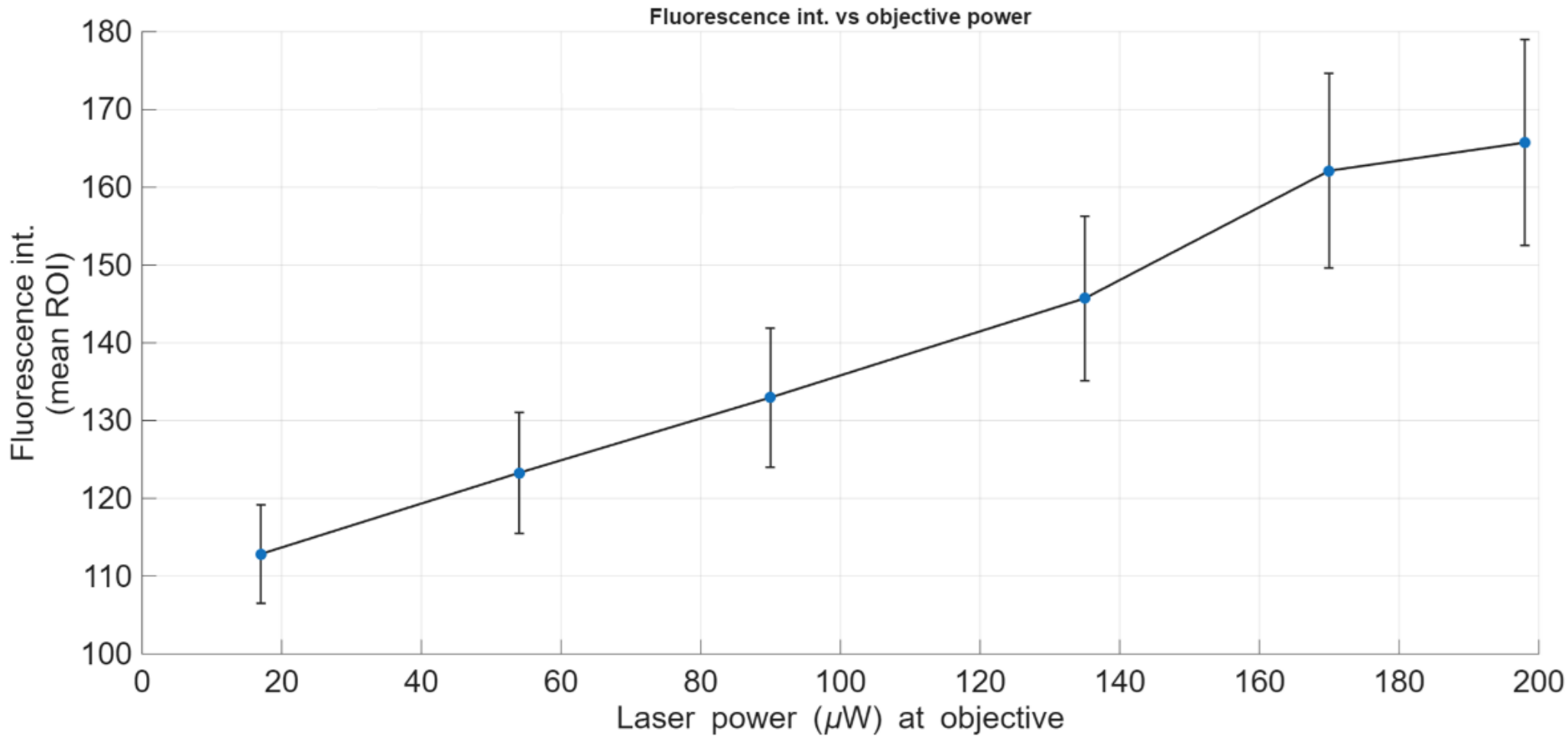

**Fig. S3.** *TIRF measurements of NBE layers are in the linear excitation range.* Integrated fluorescence vs. laser power measured at the objective. Symbols and error bars show mean ± SD area upon TIRF illumination. Overlaid line is linear fit with the data, indicating that even for highest powers used we are still not saturating the fluorophores.

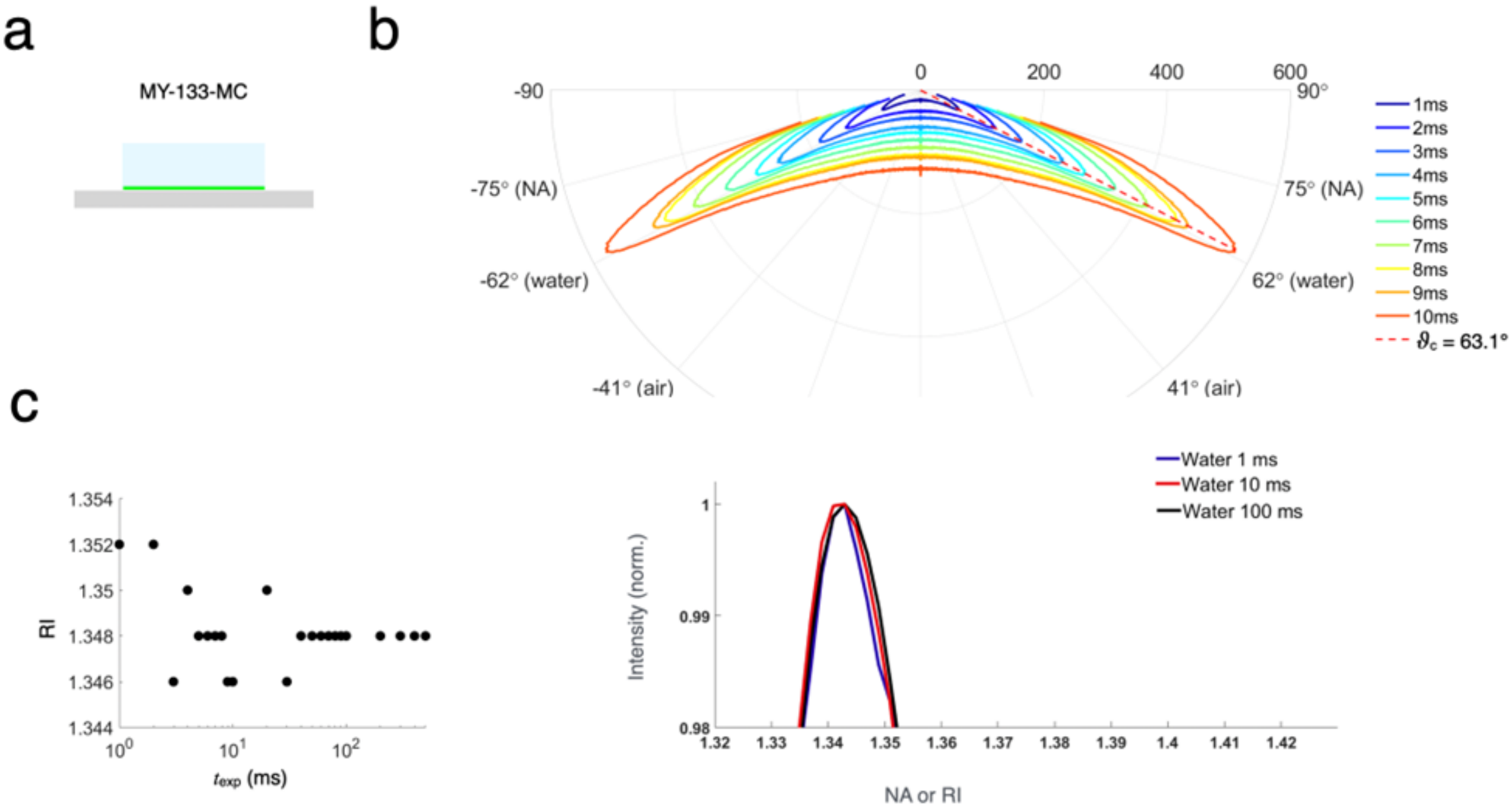

**Fig. S4.** *Measured radiation patterns and RI-values are robust measures down to exposure times of a few milliseconds.* (a), Schematic representation of the sample used in this experiment. A NBE-ATTO488 coated coverslip was overlaid with a several µm-thick layer of MY-133-MC polymer and BFP images acquired at different exposure times upon 488-nm TIRF excitation. Images were analysed as described, by first finding the center with SAM and extracting the azimuthally averaged radial intensity line profile. (b), *top*, polar-plot representation of the measured radiation patterns as a function of exposure time between 1 and 10 ms (pseudo-colour coded *blue* to *red*). *Bottom,* cropped SAF peaks from the top panel vs. RI (or, equivalently, NA) show a consistent peak position. Curves were smoothed with a Savitzky-Golay algorithm (5th-order polynomial, frame of 31). (c), measured RI vs. exposure time reveals an asymptote of RI = 1.348. The discrete spacing of the measure RI values results from the pixelation of the BFP image and hence intensity profile. We did no further efforts to fit the precise peak position, as the used data is already azimuthally averaged (see Fig. S3) and smoothed.

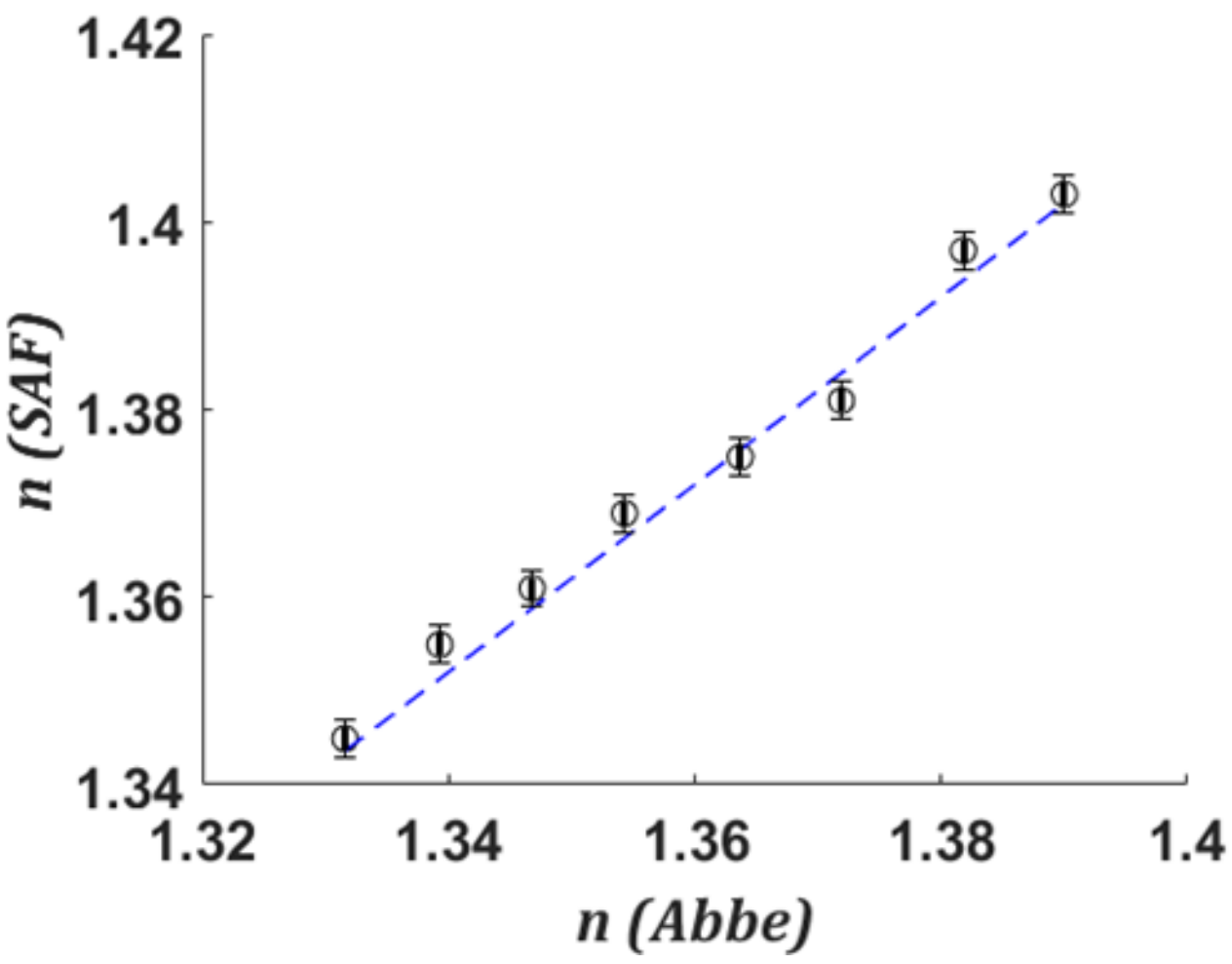

**Fig. S5.** *Comparison of SAF-based RI measurements and Abbe refractometry*. Drops of the same Brix solutions were measured on our microscope, shown as *n*(SAF), and on a conventional Abbe refractometer, *n*(Abbe), respectively. *Blue dashed* line is a linear fit with the data, having a slope of 1 and 0.013 RI units offset. Error bars reflect the accuracy of finding $\vartheta_c$ on radially averaged intensity profiles, and are limited by, (*i*), the accuracy of finding the centers of the BFP image, and (*ii*) of determining the peak position on the azimuthally averaged intensity line profiles (see Fig.S4). Note the increased precision and smaller offset compared to our earlier work (see ref. [2]) resulting from the much brighter, more uniform and thinner NBE layers compared to the then used small-chemical fluorophore layers.

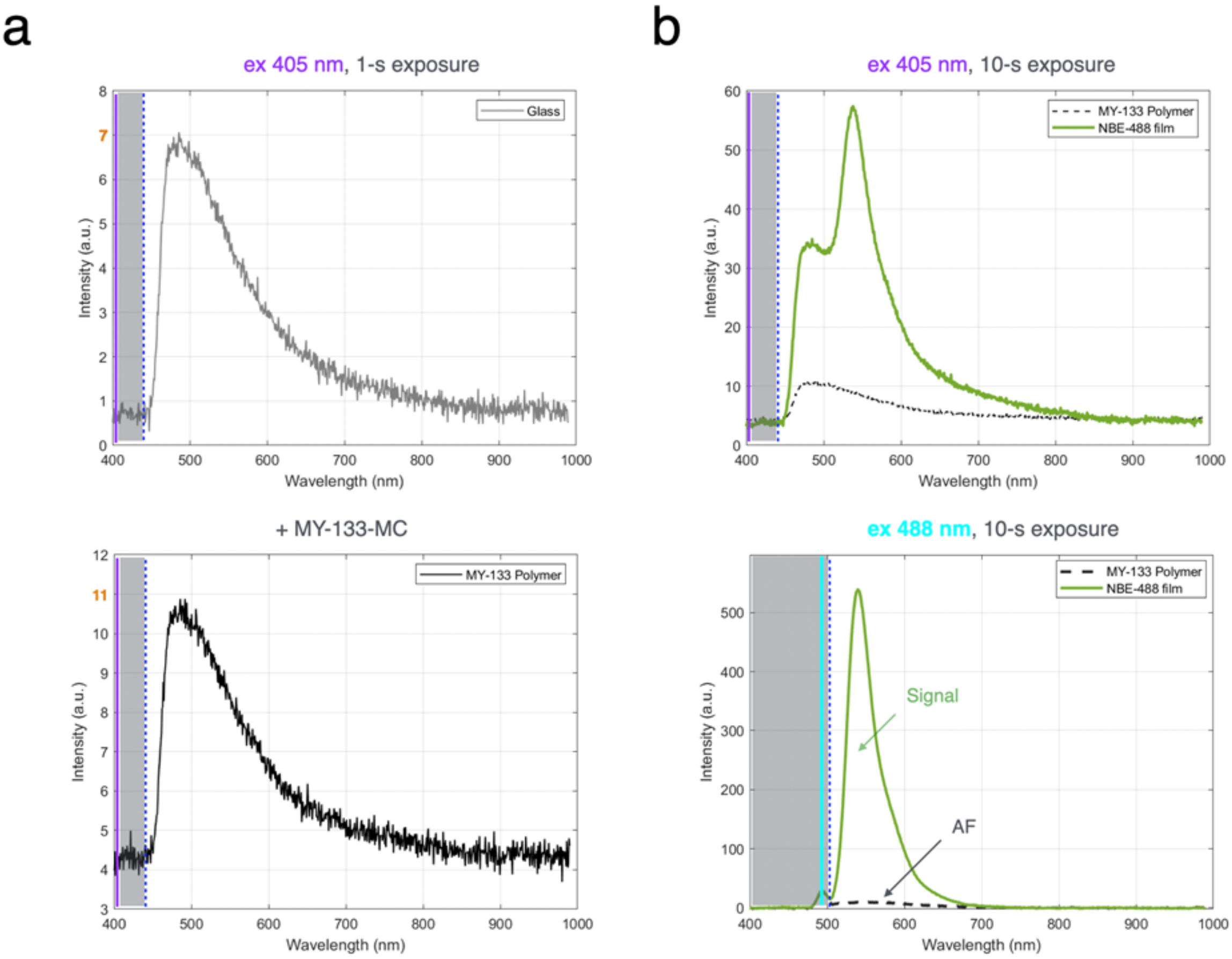

**Fig. S6.** *MY-133-MC polymer has a negligible autofluorescence relative to NBE layers.* (a), average ($n$ = 10) emission spectrum upon 405-nm excitation of a bare BK-7 glass coverslip (*top*), and the same, topped with a MY-133-MC polymer drop (*bottom*). $t_{exp}$ = 1 s for both. Note the near-identical shape and only slightly increased intensity (*red*). (b), *top*, same, upon 405-nm excitation. *Bottom*, same, upon 488-nm excitation as used in our experiments. Non-specific fluorescence is less than 4% of the NBE-ATTO488 signal. $t_{exp}$ = 10 s for both. See Fig. S11 for the filters used and Table S2 for a quantification of the measured intensities.

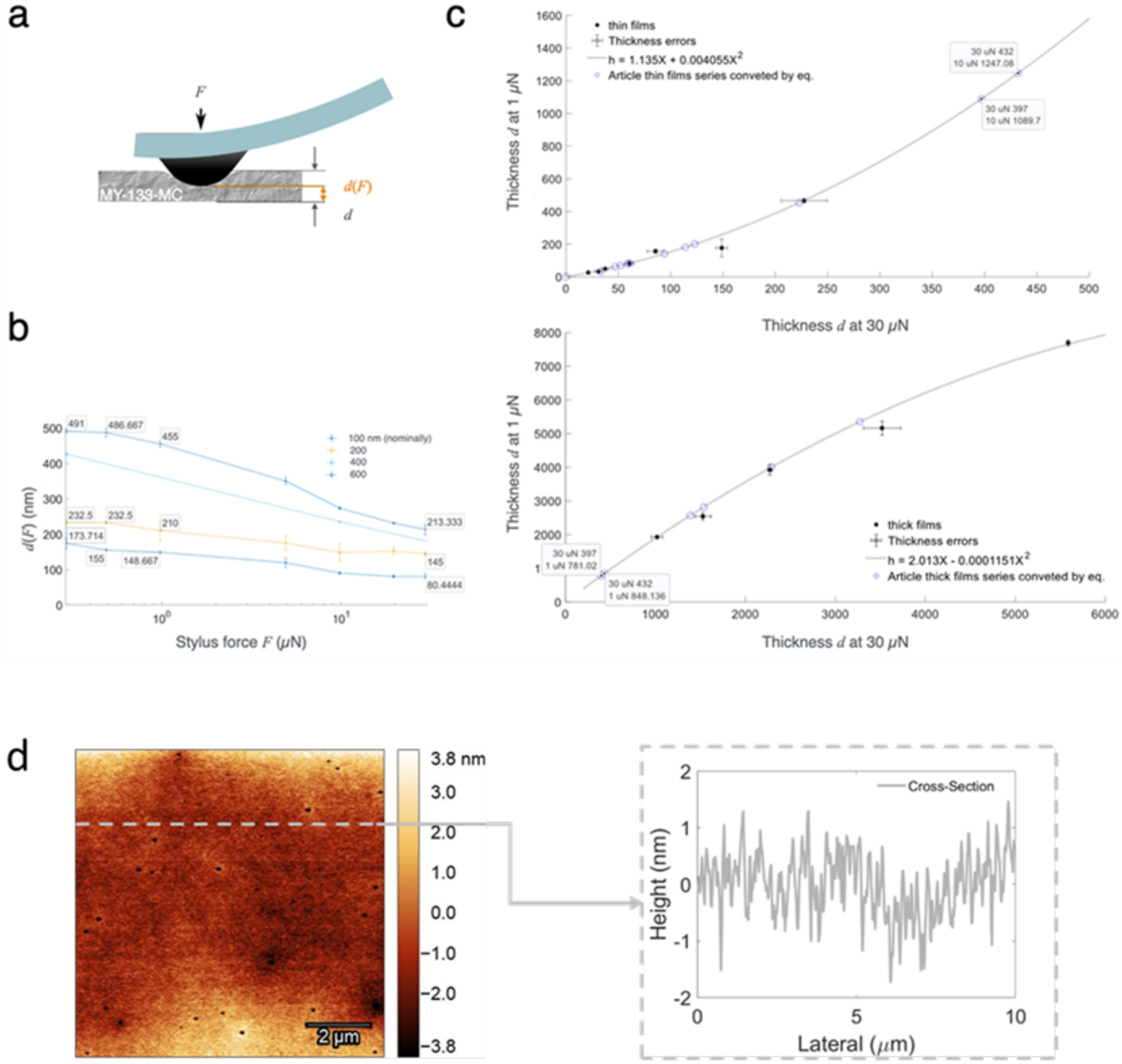

**Fig. S7.** *Measured thicknesses of the MY-133-MC polymer layers depended both on applied stylus force and on the thickness Δ of the prepared layers.* (a), sketch of force-induced indentation, resul-ting in a force-dependent apparent thickness *d*(*F*), which is not unexpected for soft matter[3,4]. (b), semi-log plot of the measured apparent thickness *d* against the applied force *F* revealed an almost mono-exponential behavior for small thicknesses and a more pronounced indentation (steeper slope) for the thicker samples. We fitted - for each produced layer - a falling exponential with the measured *d*(*F*) and arbitrarily *defined* the intersection of the fit with 1 µN force as its *effective* thickness. While this obviously represents a choice, we are at the low-force end of the distribution, and the applied stylus force of 1 µN is in the range of published values for polymer layers (Tzukrug & Singamaneni (2011)[5]). Positive and negative 'error' bars were generated from the values measured at 0.5 and 1.5 µN, respectively, and give an idea of the variation with force in that range. (c), parametric plot of apparent thickness *d*(1µN) vs. *d*(30µN) reveals different regimes, depending on the layer thickness: for thin layers, surface effects dominate and lead to an increasing stiffness for thinner layers where as volume effects and long-range order dominate for µm-thick layers. (d) AFM topography image and corresponding height profile along the grey dashed line. Image analysis was performed using Gwyddion software. The image was auto-leveled, and the central point of the color scale was set to zero. The height profile, extracted from one line (512 × 512 pixels image), was similarly normalized by setting the mean intensity value to zero.

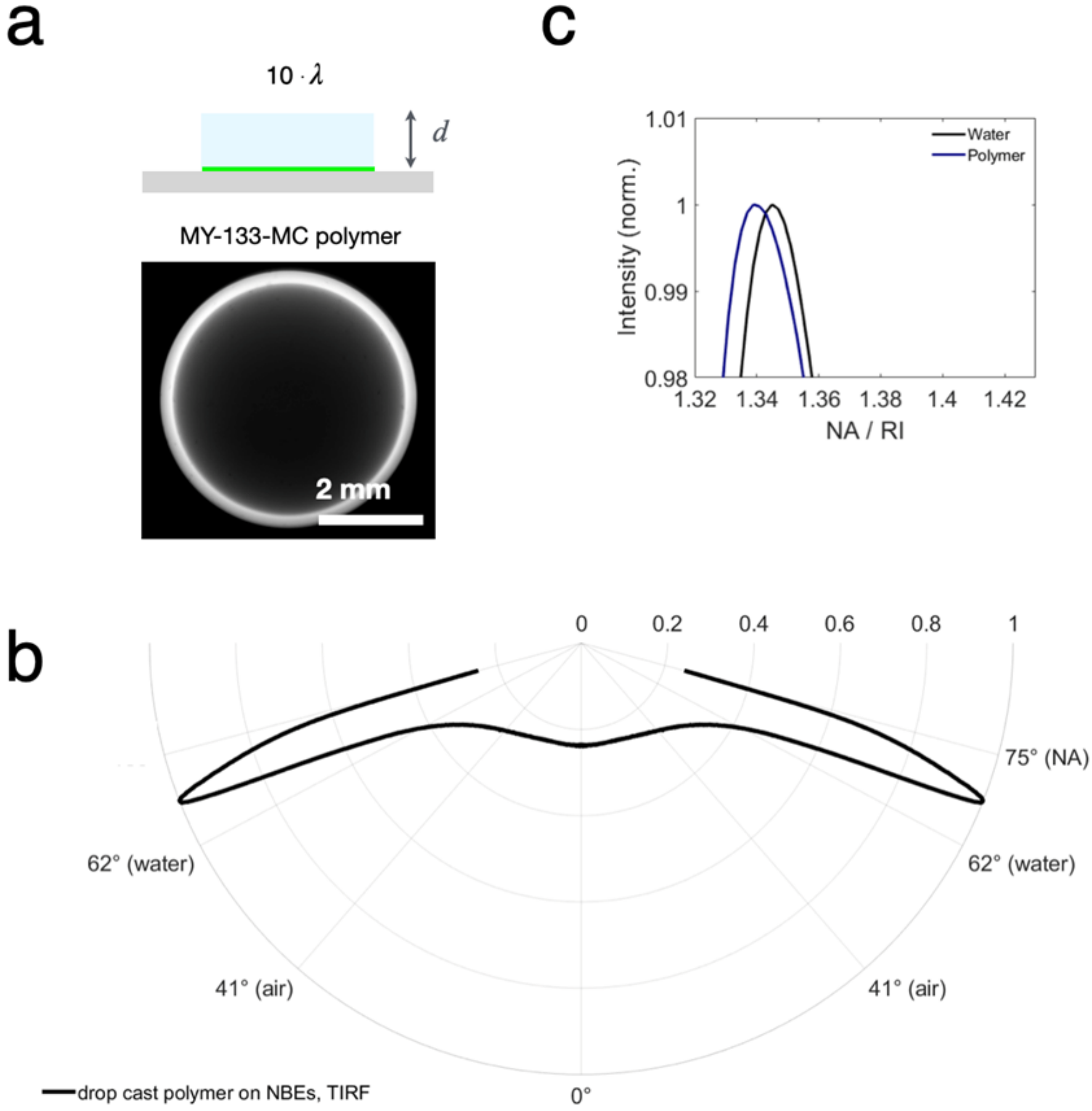

**Fig. S8.** *Drop-cast polymer films have the same refractive index as the the bulk polymer*. (a), *top*, illustration of the used NBE-ATTO488 (*green*) covered smart coverslips, coated with a µm-thick MY-133-MC polymer layer (*blue*). *Bottom*, corresponding BFP image. (b), azimuthally averaged radiation pattern of a µm-thick MY-133-MC film showed a SAF emission peak at 1.339 ± 0.002 (*black*). The peak for water is shown for comparison, $n$ = 1.345 ± 0.002, (c, *blue*).

a

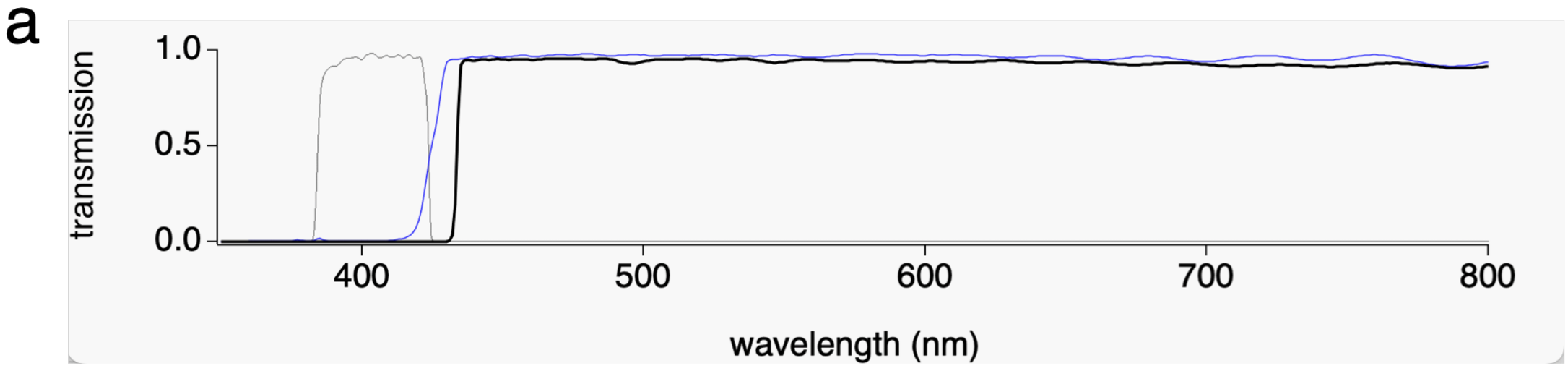

b

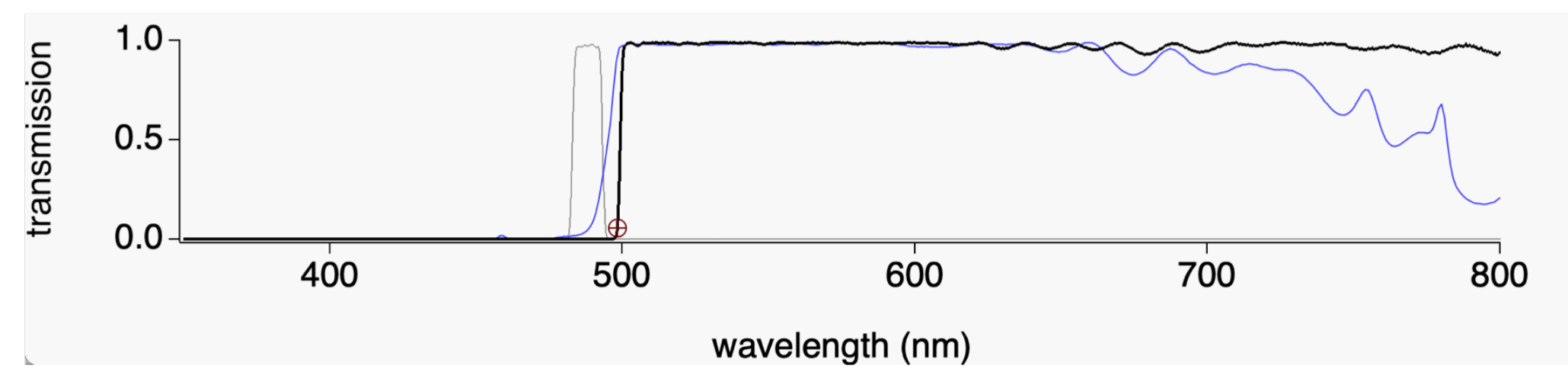

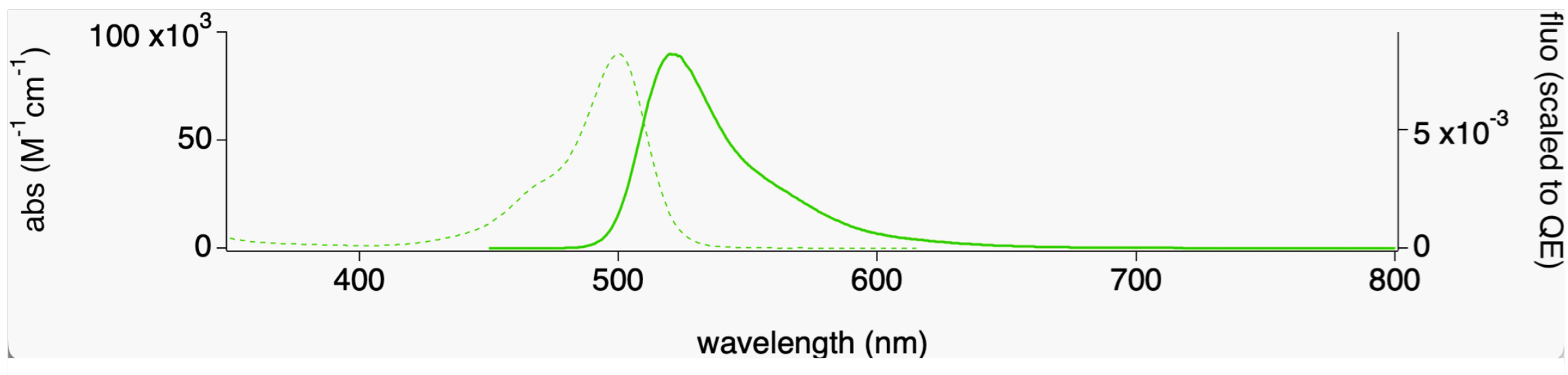

c

**Fig. S9.** *Instrument and reference spectra for (auto-) fluorescence micro-spectrometry.* (a), trans-mission spectra of the used excitation (*grey*), dichroic (*blue*) and emission filters (*black*) used for 405-nm excitation (which favours autofluorescence, AF). (b), same, for 488-nm excitation. (c), fluo-rescence excitation (*dashed*) and emission spectrum (*through* line) for ATTO488 dye. Spectra from Chroma Technology (a, b) and ATTO-tec (c), respectively.

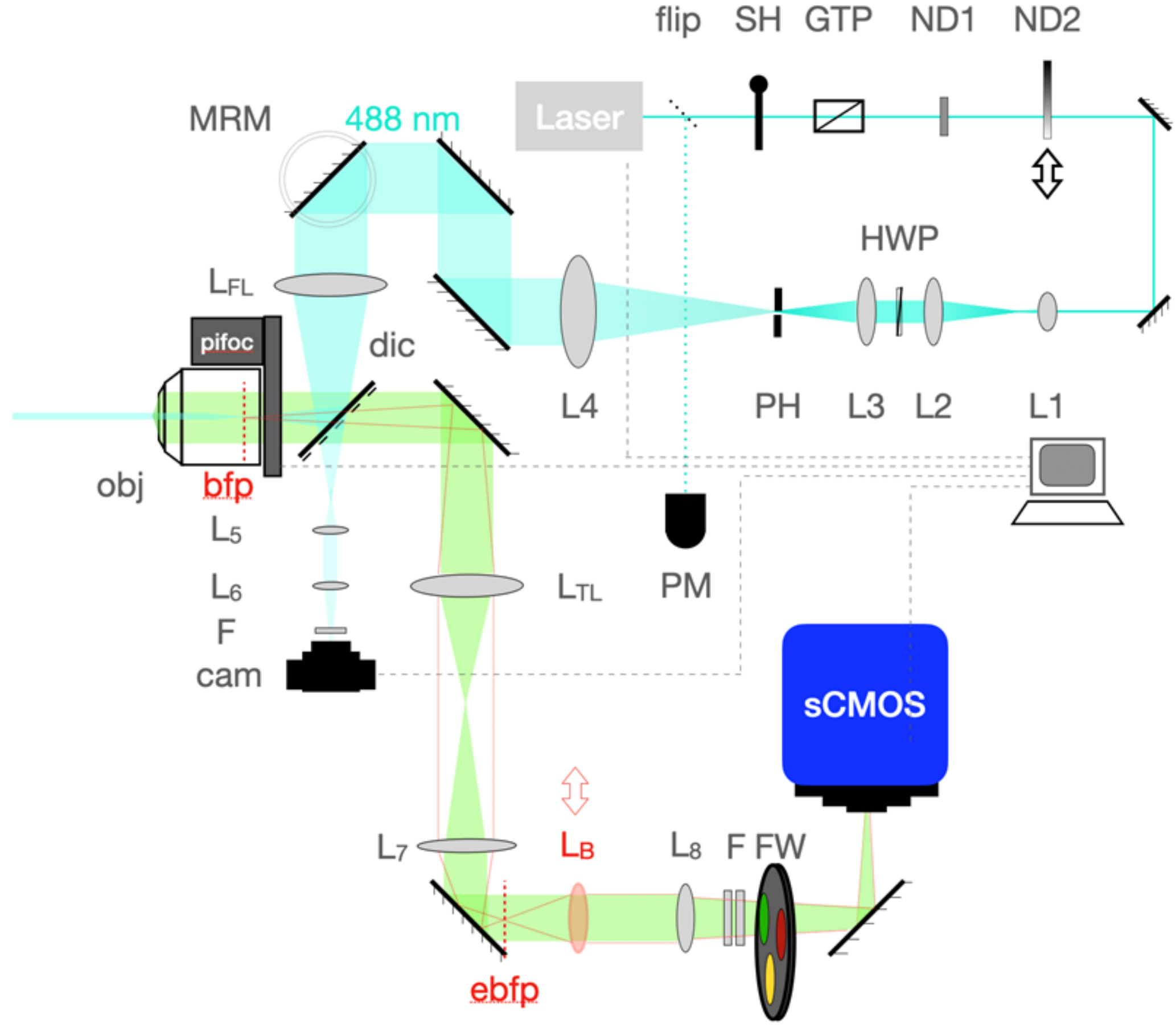

**Fig. S10** *Custom TIR-SAF super-resolution microscope*. Simplified optical layout of our custom inverted microscope assembled from optical-bench components : flip - flippable mirror, SH - shutter, GTP - Glan-Thompson polarizer, ND - neutral-density filter, L - lens, PH - pinhole, MRM - mechanical rotation mount, $L_{FL}$ - focusing lens, dic - dichroic mirror, pifoc - piezo-electric focus drive, obj - microscope objective, $L_{TL}$ - tube lens, (e)bfp - (equivalent) back-focal plane, $L_B$ - motorized Bertrand lens, F - filters, FW - filter wheel, sCMOS - scientific complementary metal-oxide sensor (camera). *turquoise* and *green* shading indicate, respectively, excitation and emission optical paths. cam is a small CCD camera for beam diagnosis that uses the residual transmission of the dichroic. Red dashed lines illustrate effect of the phase telescope when the Bertrand lens is in place. Grey dash schematises computer control via our custom graphical user interface (GUI). Also see refs. [6] and [7] for details on the setup and GUI, respectively.

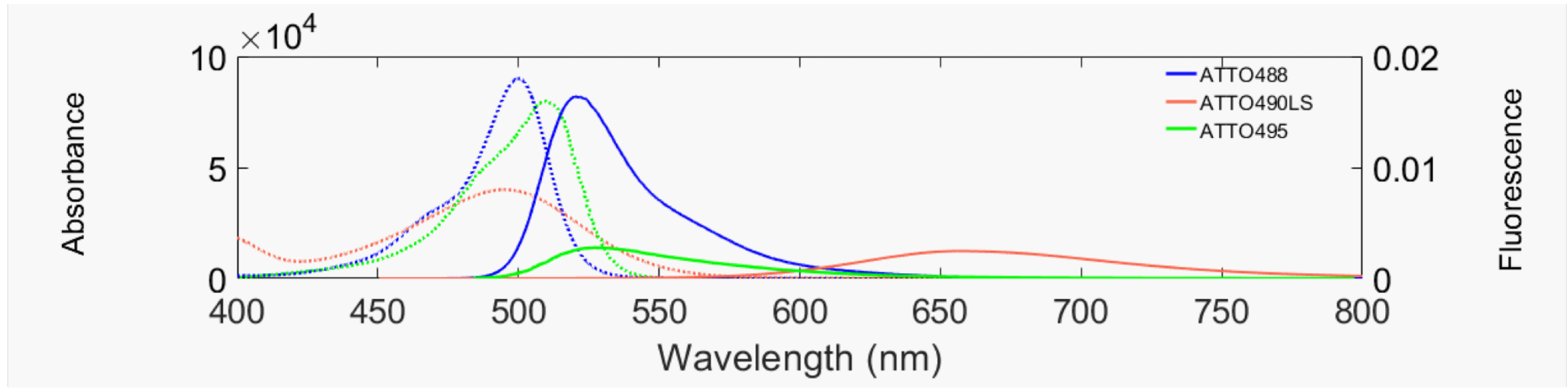

a

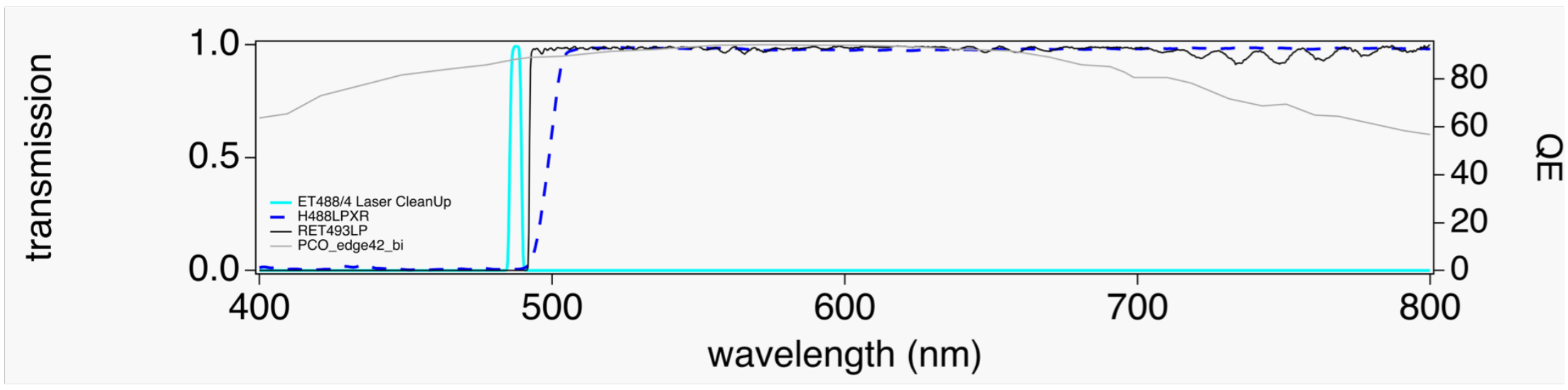

b

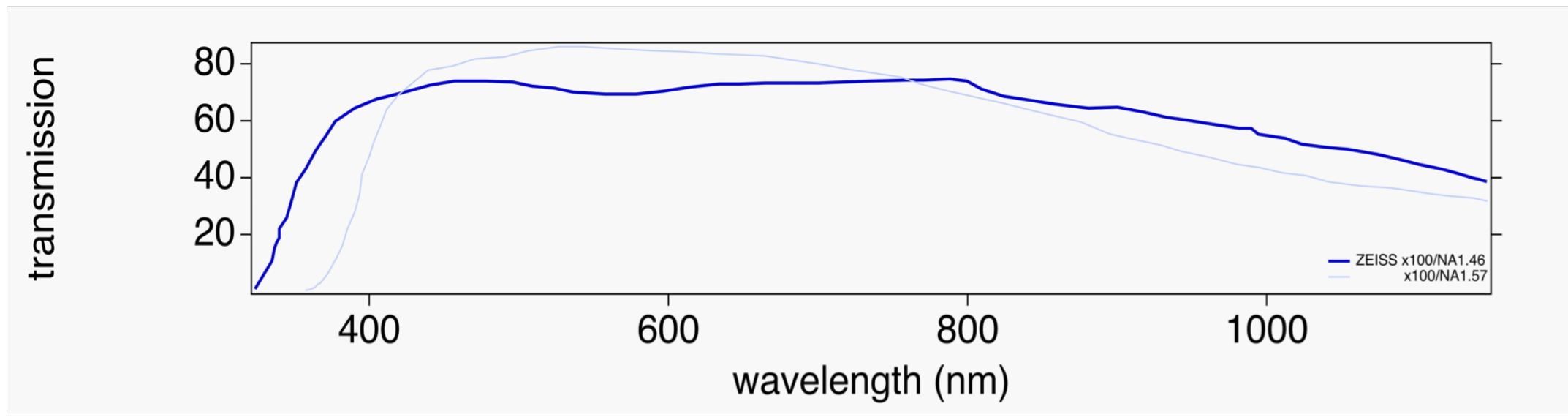

c

**Fig. S11.** *Source, dye, filter and camera spectra*. (a), *dashed*, fluorescence excitation and emission spectra, *solid*, of ATTO488 (*blue*), ATTO490LS dyes (*red*) and ATTO495 (*green*), respectively. Excitation spectra are scaled to peak molar extinction (abs), emission spectra (fluo) are scaled so that the area under the curve is equivalent to the fluorescence quantum efficiency (QE), to allow for a direct intensity comparison. (b), transmission spectra of the laser clean-up (*turquoise*), dichroic (*blue dashed*), and emission filter (*black*), as well as camera sensitivity (*grey*), respectively. Spectral data from the suppliers' websites: ATTO-tec for the dyes, AHF-Analysentechnik for the filters, and Excelitas-PCO for the CMOS camera, respectively. (c), transmission curves for the ZEISS NA-1.46 objective, *blue*, and the NA-1.57 lens, *light blue*. Where not available in numerical form (ASCII), spectra were canned from manufacturer data with WebPlotDigitizer v4 (automeris.io).

**Tables**

**Table S1.** ***Measured excitation and fluorescence emission peak wavelengths***

***for native dyes, NBE solutions and NBE layers***

| | ***NBE (dye) Solutions*** | | ***NBE layers*** | |
|---|---|---|---|---|
| **(Native dye)** | $\lambda_{ex}^{(max)}$ **(nm)** | $\lambda_{em}^{(max)}$ **(nm)** | $\lambda_{ex}^{(max)}$ **(nm)** | $\lambda_{em}^{(max)}$ **(nm)** |
| **ATTO425** | **432**<br>**(439)** | **480**<br>**(485)** | **435** | **484** |
| **ATTO488** | **498**<br>**(500)** | **534**<br>**(523)** | **501** | **527** |
| **ATTO490LS** | **490**<br>**(496)** | **590**<br>**(661)** | **493** | **640** |

[a] from Olevsko *et al.* (2025), Ref. [1]

*N.A.* - not available

**Table S2.** ***Measured fluorescence emission upon 405- and 488-nm excitation for bare glass, thick MY-133-MC drop and NBE-ATTO488 layer***

| | $\lambda_{ex}$ = ***405 nm*** [b] | | ***488 nm*** | |
|---|---|---|---|---|
| **(Native dye)** | $\lambda_{em}^{(max)}$ **(nm)** | $F_{tot}$ [c] **(% of NBE)** | $\lambda_{em}^{(max)}$ **(nm)** | $F_{tot}$ [c] **(% of NBE)** |
| **BK-7 glass** | **432** | **8.7** | ***N.A.*** | **-** |
| **MY-133-MC drop** | **480** | **13.3** | **539** | **3.8** |
| **NBE-ATTO488 layer** [a] | **527** | **(100)** | **527** | **(100)** |

[a] 8-months old uncoated NBE layer, representing a worst-case

[b] off-peak excitation, NBE-ATTO488 $\lambda_{ex}^{(max)}$ = 498 nm

[c] integrated fluorescence above baseline (from an average of 10 spectra taken at 10-s exposure time)

**Movie S1 (separate file).** *Time-lapse BFP imaging of solvent (HFE-7500) evaporation*. Time-lapse of back focal plane (BFP) images acquired by our SAF-TIRF microscope during evaporation of a 2-µl hydrofluoroether (HFE-7500) droplet on a $SiO_2$-protected smart surface (NBEs labeled with ATTO488, ~15 nm $SiO_2$ overlayer). The movie shows the raw experimental observable underlying Fig. 4, illustrating the temporal evolution of the SAF radiation pattern as the local dielectric environment transitions from liquid HFE to air. Images were recorded at 50 frames per second and uniformly auto-contrasted using FIJI (ImageJ) with identical contrast limits applied to all images to preserve relative intensity changes.

**Movie S2 (separate file).** *Time-resolved radial intensity profiles extracted from BFP images*. Movie showing the temporal evolution of radial line-intensity profiles extracted from the BFP image time series shown in Supporting Video S1. A straight radial cross-section was taken at an azimuthal angle of 45° with respect to the image x-axis, originating from the center of the BFP image (center position determined independently using SAM analysis in MATLAB). The FIJI (ImageJ) live line-profile tool was used to visualize intensity as a function of radial position for each frame in the time-lapse sequence. The dynamic evolution of the intensity profiles was recorded by screen capture using Free Cam 8 software while scrolling through the image stack. The resulting movie was accelerated by a factor of 3.5 to match the duration of Video S1.

**SI References**